\documentclass[final,5p,times,twocolumn]{elsarticle}

\makeatletter
\def\ps@pprintTitle{%
 \let\@oddhead\@empty
 \let\@evenhead\@empty
 \def\@oddfoot{}%
 \let\@evenfoot\@oddfoot}
\makeatother

\usepackage{graphicx}
\usepackage{xfrac}
\usepackage{amsmath}
\usepackage{mathrsfs}
\usepackage{amsfonts}
\usepackage{lineno}
\usepackage{hyphenat}
\usepackage{cases}
\usepackage {mathtools}
\usepackage {bm}
\usepackage{amsthm}
\usepackage{amssymb}
\usepackage[english]{babel}
\usepackage{lipsum}
\usepackage{float}
\usepackage{nicefrac}
\usepackage{textcomp}
\usepackage{eufrak}
\usepackage{bbold}
\usepackage{cuted}
\usepackage{widetext}
\usepackage[colorlinks,citecolor=blue!50!black,linkcolor=blue!50!black,urlcolor=blue!50!black]{hyperref}
\usepackage{cleveref} 

\DeclareMathOperator{\sech}{sech}
\newcommand{\Var}{\operatorname{Var}}

\newtheorem{remark}{remark}

\journal{Journal of Magnetic Resonance}

\begin{document}

\begin{frontmatter}




\title{Spin dynamics during chirped pulses: applications to homonuclear decoupling and broadband excitation}



\author{Mohammadali Foroozandeh\corref{M. Foroozandeh}}

\address{Chemistry Research Laboratory, University of Oxford, Mansfield Road, Oxford OX1 3TA, UK}

\ead{mohammadali.foroozandeh@chem.ox.ac.uk}

\begin{abstract}

Swept-frequency pulses have found applications in a wide range of areas including spectroscopic techniques where efficient control of spins is required. For many of these applications, a good understanding of the evolution of spin systems during these pulses plays a vital role, not only in describing the mechanism of techniques, but also in enabling new methodologies. In magnetic resonance spectroscopy, broadband inversion, refocusing, and excitation using these pulses are among the most used applications in NMR, ESR, MRI, and $in$ $vivo$ MRS. In the present survey, a general expression for chirped pulses will be introduced, and some numerical approaches to calculate the spin dynamics during chirped pulses via solutions of the well-known Liouville–von Neumann equation and the lesser-explored Wei-Norman Lie algebra along with comprehensive examples are presented. In both cases, spin state trajectories are calculated using the solution of differential equations. Additionally, applications of the proposed methods to study the spin dynamics during the PSYCHE pulse element for broadband homonuclear decoupling and the CHORUS sequence for broadband excitation will be presented.

\end{abstract}


%
%
%
%

\begin{keyword}

Chirped pulses \sep Liouville–von Neumann equation \sep Wei-Norman Lie algebra \sep differential equations \sep homonuclear decoupling \sep PSYCHE \sep broadband excitation \sep CHORUS

\end{keyword}

\end{frontmatter}


\section{Introduction}
\label{sec:intro}

Swept-frequency pulses are a group of parametric pulses during which the frequency of irradiation varies with time. They are generally called according to the distribution function of their frequency sweep, e.g. linear (chirped pulses), hyperbolic secant (HS pulses), hyperbolic tangent (Tanh pulses), etc. In particular the ability of swept-frequency pulses to satisfy adiabatic condition \cite{Kato1950} under certain circumstances is very attractive in many applications \cite{Tannus1997,Garwood2001,Tesiram2002,Meriles2003}. The adiabatic properties of swept-frequency pulses cause that sometimes, misleadingly, these pulses are called "adiabatic pulses" even when they do not satisfy the adiabatic condition. In two examples presented in this paper, chirped pulses are used partially or totally below the adiabatic threshold.

Swept-frequency pulses have found a surprisingly wide range of applications in magnetic resonance, including but not limited to, designing robust broadband inversion and refocusing \cite{Tycko1983,Baum1983,Tycko1984,Baum1985,Hardy1986,Bendall1987,Bohlen1990,Fujiwara1990,Ke1992,Kupce1995,Kupce1996a,Hwang1997,Hwang1998,Kupce2007,Harris2012,Spindler2013} and excitation \cite{Kunz1986,Bohlen1989,Ermakov1993a,Ermakov1993,Skinner1993,Cano2002,Power2016a,Power2016,Khaneja2017,Foroozandeh2019} pulses, designing $B_{1}$-insensitive pulses \cite{Ugurbil1987,Ugurbil1988,Merkle1992,Ke1992,Garwood1995,Graaf1996a,Graaf1996,Zijl1996,Smith2012}, broadband heteronuclear decoupling \cite{Fujiwara1993,Fu1995a,Fu1995,Kupce1995a,Kupce1996,Kupce1996b,Kupce1996c,Kupce1996a,Kupce1997,Freeman1997,Cheatham2012}, single scan NMR \cite{Andersen2005,Lin2016}, spatiotemporal encoding \cite{Dumez2013,Dumez2014,Dumez2018}, solid-state NMR \cite{Tycko1984,Kervern2007,Harris2012,Loening2012,ODell2013,Wi2019}, dynamic nuclear polarization (DNP) \cite{Kaminker2018,Can2018,Scott2018,Gao2019}, coherence suppression \cite{Titman1990,Thrippleton2003,Thrippleton2005}, broadband electron spin resonance (ESR) \cite{Forrer1996,Spindler2012,Spindler2013,Doll2014,Doll2015,Jeschke2015,Schops2015,Segawa2015,Doll2016,Keller2016,Pribitzer2016,Pribitzer2016a,Doll2017,Pribitzer2017,Bahrenberg2017,Kaminker2017,Bieber2018,Wili2018,Ritsch2019}, and designing cpmg sequences \cite{ODell2008,Hung2010,Casabianca2014,Casabianca2015}.

Additionally they have found applications in adiabatic optimal control for logic gate design for quantum computing \cite{Kuklinski1989,Gaubatz1990,Melinger1992,Broers1992,Vitanov2001,Rangelov2005,Randall2018,Zlatanov2020}, and manipulation of NV-centers in diamond \cite{Niemeyer2013,Scheuer2016,Ajoy2018,Zangara2019}.

In many of these applications, understanding of the spin dynamics during the pulse event is as important as knowing the state of the system at the end of the pulse event. This is particularly important when some trajectories for certain coherences should be enforced or suppressed. The aim of the present article is to set a unified approach for the calculation and visualisation of the spin dynamics during chirped pulses and pulse sequences consisting only of chirped pulses using the solution of a single system of ordinary differential equations (ODEs). The proposed scheme will be presented in the context of two different mathematical formalisms: the first gives access to the solution of familiar Liouville–von Neumann equation, and the second takes advantage of Wei-Norman Lie algebra, a powerful, but rarely explored technique in magnetic resonance.

The motivation behind this proposition is threefold: (i) As opposed to the conventional density matrix approach, relying on matrix exponentiation and piecewise constant propagation, ODE integrators have an adaptive time-step selection which changes depends on the local frequency of the oscillation. This property enables much faster computation of the spin dynamics during swept-frequency pulses, with fewer time steps for the same accuracy. This feature is especially attractive when computation of the spin dynamics during swept-frequency pulses is part of an iterative process like optimisation. (ii) Although the conventional density matrix approach could be more general, as it can, in principle, handle sudden changes and discontinuities in the Hamiltonian, swept-frequency pulses are smooth, time-continuous by nature, ideal for adaptive non-stiff ODE integrators (e.g. ode45 in MATLAB). (iii) both proposed approaches based on Liouville-von Neumann equation and Wei-Norman Lie algebra can lead to analytical or closed-form solutions and hold promise for more efficient construction and optimisation of pulse sequences consisting of swept frequency pulses. 

This paper is structured as follows: in \cref{sec:Theory} a general expression for swept-frequency (chirped pulses in particular) will be introduced, followed by detailed presentations of Liouville–von Neumann equation and Wei-Norman Lie algebra. For both approaches, general mathematical formalisms are presented, along with examples representing the applications of these formalisms in cases relevant to NMR. Finally, each method has a "demo" NMR technique, demonstrating how the proposed approach gives access to the desired information about the spin dynamics. In \cref{sec:applic} a comprehensive usage of the proposed approaches will be presented for broadband homonuclear decoupling using a PSYCHE pulse element and broadband chirped excitation using a CHORUS pulse sequence.

\section{Theory}
\label{sec:Theory}

\begin{remark}
All solutions presented in this article are numeric, obtained using the adaptive Runge-Kutta method \cite{Dormand1980}. Similar results can be obtained using other proposed methods relying on the approximation of matrix exponentials and Fokker-Planck formalism \cite{Kuprov2016,Allami2019}, available via SPINACH package \cite{Hogben2011}. Approximated analytical solutions can be obtained using instantaneous flip approximation during chirped pulses \cite{Dumez2018}. Exact and explicit closed-form or analytical solutions can be obtained using various techniques, including the application of integrable systems \cite{Sinitsyn2017,Sinitsyn2018} and algebraic graph theory \cite{Giscard2015}. These solutions, even for coupled spins, can be represented as analytical or closed-form expressions using transcendental functions (hypergeometric, Whittaker, Fresnel integrals, etc.) and will be presented elsewhere. Although, some analytical solutions of the Bloch equations for swept-frequency pulses, in particular hyperbolic secant pulses, have been presented in the literature \cite{Hioe1984,Zhang2017}.
\end{remark}

\subsection{Generalised chirped pulse}

It is typical to write the general form of swept-frequency pulses as:

\begin{equation}
S(t) = \omega_{1}(t) \exp (\mathrm{i}\phi(t))
\end{equation}

where the amplitude envelope $\omega_{1}(t)$ and the phase $\phi(t)$ are continuous, real-valued functions of time. The relationship between frequency sweep function $\omega(t)$ and phase $\phi(t)$ of the pulse can be written as:

\begin{equation}
\phi(t) = \int_{0}^{\tau_p} \omega(t) dt
\end{equation}

Although in the rest of this paper only chirped pulses with a linear frequency sweep, and therefore a quadratic phase, are considered, the framework is general and can be applied to any form of parametrised pulses with time-continuous parameters.

We start with the introduction of the most general form of a chirped pulse. As opposed to the conventional 3-parameter chirp pulse described by amplitude ($\omega_{1}$), duration ($\tau_{p}$), and bandwidth ($\Delta F$), these pulses are described by 6 parameters: amplitude ($\omega_{1}$), bandwidth ($\Delta F$), duration ($\tau_{p}$), overall phase ($\phi_{0}$), time offset ($\delta_{t}$), and frequency offset ($\delta_{f}$). These additional parameterisations, as will be shown later, allow us to construct any type of pulse sequence consisting of chirped pulses, either concatenated or superimposed, very easily as a single sum of time-continuous functions. Additionally, any further offset modulation or overall phase variation (as in phase cycling) can be incorporated in this expression.

Most commonly, smoothing of the time-envelope for chirped pulses is achieved via the application of WURST \cite{Kupce1995a}, or quarter-sine \cite{Bohlen1993} smoothing functions. For a generalised chirped pulse, presented here, the time-envelope is a super-Gaussian distribution, covering the whole sequence:

\begin{equation}
G(t)=\exp \left[-2^{n+2} \left(\frac{t-\delta_{t}}{\tau_{p}}\right)^{n} \right], \qquad n \in 2 \mathbb{Z}^{+}
\end{equation}

Here $\delta_{t}$ and $\tau_{p}$ are related to the mean and variance of the distribution respectively, and $n$ is an even number determining the smoothing of the time envelope, where for $n=2$ a normal Gaussian envelope will be obtained. Generally $n$ in the range of 20 to 40 will be adequate for most applications.  ,.

The complete expression for a generalised chirped pulse can be written as:

\begin{equation}
\label{eq:suchi}
\begin{aligned}
S(t)=\omega_{1} \exp &\left[-2^{n+2} \left(\frac{t-\delta_{t}}{\tau_{p}}\right)^{n} \right.\\
& \left. +\mathrm{i} \left(\phi_{0}+\frac{\pi \Delta F (t-\delta_{t})^{2}}{\tau_{p}}-2 \pi \delta_{f} (t-\delta_{t})\right)\right]
\end{aligned}
\end{equation}

\Cref{eq:suchi} can also be represented in Cartesian coordinates as:

\begin{equation}
S(t)=C_{x}(t) + \mathrm{i} C_{y}(t)
\end{equation}

where

\begin{equation}
\label{eq:genchirpx1}
\begin{split}
C_{x}(t)=&\omega_{1} \exp \left[-2^{n+2} \left(\frac{t-\delta_{t}}{\tau_{p}}\right)^{n} \right] \\
&\cos\left[\phi_{0}+\frac{\pi \Delta F (t-\delta_{t})^{2}}{\tau_{p}}-2 \pi \delta_{f} (t-\delta_{t})\right]
\end{split}
\end{equation}

and

\begin{equation}
\label{eq:genchirpy1}
\begin{split}
C_{y}(t)=&\omega_{1} \exp \left[-2^{n+2} \left(\frac{t-\delta_{t}}{\tau_{p}}\right)^{n} \right] \\
&\sin\left[\phi_{0}+\frac{\pi \Delta F (t-\delta_{t})^{2}}{\tau_{p}}-2 \pi \delta_{f} (t-\delta_{t})\right]
\end{split}
\end{equation}

\begin{figure}
  \includegraphics[width=\linewidth]{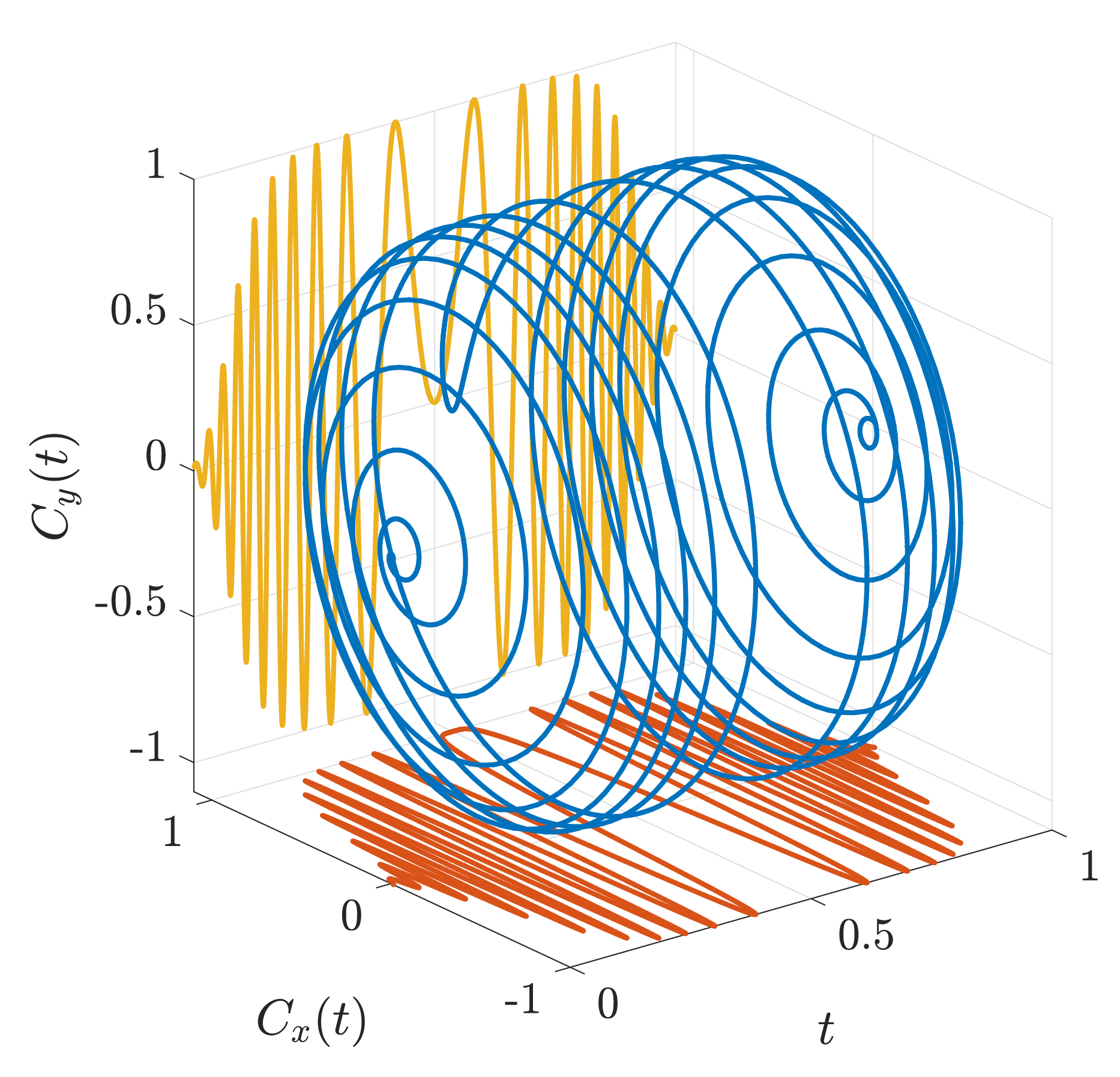}\\
  \caption{Graphical representation of a complex chirped pulse as presented in \cref{eq:suchi}, and its orthogonal components, $C_{x}(t)$ and $C_{y}(t)$, for a pulse with $\Delta F = 60$ kHz, $\tau_p = 1$ ms, $\delta_t = 0.5$ ms, $\phi_{0} = 0$, $\delta_f = 0$, $\omega_{1}=1$, and $n = 10$.}
  \label{fig:fig_1}
\end{figure}

\Cref{fig:fig_1} shows a complex chirped pulse with its real ($C_{x}(t)$) and imaginary ($C_{y}(t)$) components.

\Cref{eq:suchi,eq:genchirpx1,eq:genchirpy1} can be used to construct any pulse sequence consisting of chirped pulses as a sum of generalised chirped pulses:

\begin{equation}
\label{eq:genchirp}
\begin{split}
S(t)=\sum_{i=1}^{I}\omega_{1}^{(i)} &\exp \left[-2^{n+2} \left(\frac{t-\delta_{t}^{(i)}}{\tau_{p}^{(i)}}\right)^{n} \right. \\
&\left. + \mathrm{i} \left(\phi_{0}^{(i)}+\frac{\pi \Delta F (t-\delta_{t}^{(i)})^{2}}{\tau_{p}^{(i)}}-2 \pi \delta_{f}^{(i)} (t-\delta_{t}^{(i)})\right)\right]
\end{split}
\end{equation}

Where $I$ is the number of chirped pulses. Note that due to the nature of super-Gaussian time envelopes, there is no need for explicit inclusion of delays in the sequence and those can be incorporated using an appropriate time offset $\delta_{t}$ of each pulse. Again the Cartesian components of $S(t)$ are:

\begin{equation}
\label{eq:genchirpx}
\begin{split}
C_{x}(t)=\sum_{i=1}^{I} & \omega_{1}^{(i)} \exp \left[-2^{n+2} \left(\frac{t-\delta_{t}^{(i)}}{\tau_{p}^{(i)}}\right)^{n} \right] \\
&\cos\left[\phi_{0}^{(i)}+\frac{\pi \Delta F (t-\delta_{t}^{(i)})^{2}}{\tau_{p}^{(i)}}-2 \pi \delta_{f}^{(i)} (t-\delta_{t}^{(i)})\right]
\end{split}
\end{equation}

and

\begin{equation}
\label{eq:genchirpy}
\begin{split}
C_{y}(t)=\sum_{i=1}^{I} & \omega_{1}^{(i)} \exp \left[-2^{n+2} \left(\frac{t-\delta_{t}^{(i)}}{\tau_{p}^{(i)}}\right)^{n} \right] \\
&\sin\left[\phi_{0}^{(i)}+\frac{\pi \Delta F (t-\delta_{t}^{(i)})^{2}}{\tau_{p}^{(i)}}-2 \pi \delta_{f}^{(i)} (t-\delta_{t}^{(i)})\right]
\end{split}
\end{equation}

\begin{figure}
  \includegraphics[width=\linewidth]{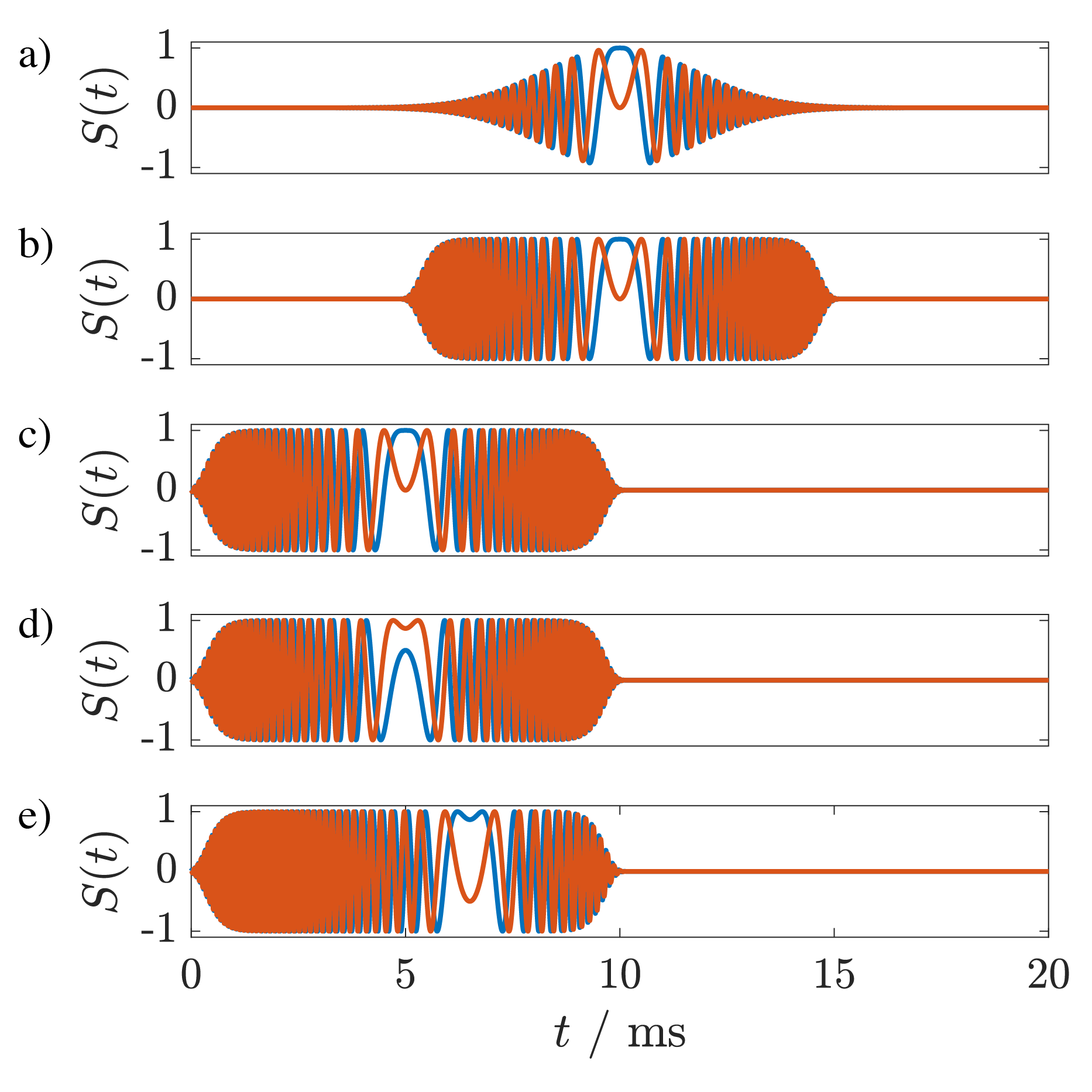}\\
  \caption{Graphical representation of Cartesian components ($C_{x}(t)$ in blue and $C_{y}(t)$ in red) of a chirped pulse with $\Delta F = 20$ kHz, $\tau_p = 10$ ms, and $\omega_{1}=1$ in a time frame of 20 ms; a) with $\phi_{0} = 0$, $\delta_t = 10$ ms, $\delta_f = 0$, and $n = 2$, b) the same as (a) but with $n = 20$, c) the same as (b) but with $\delta_t = 5$ ms, d) the same as (c) but with $\phi_{0} = \nicefrac{\pi}{3}$, and e) the same as (d) but with $\delta_f = 3000$ Hz.}
  \label{fig:fig_2}
\end{figure}

\Cref{fig:fig_2} shows the effect of super-Gaussian smoothing and additional parametrisation on a chirped waveform.

As will be demonstrated in \cref{subsec:chorus}, the proposed general scheme for defining chirped pulses is not only useful for the generation and description of chirped pulse sequences, but also when certain optimisation approaches take advantage of parametrised, or semi-parametrised waveforms \cite{Meister2014,Goodwin2018}.


\subsection{Liouville–von Neumann equation} 

In this section we consider construction of a matrix homogeneous ODE (ordinary differential equation) to have access to time-dependent evolution of each member of the basis set $\mathcal{L}$. Explicit examples will be provided in the succeeding sections, but first let us consider the most general case. For a basis set $\mathcal{L}=\left\{X_{1}, X_{2}, \ldots, X_{N}\right\}$, we consider that the set is closed under all possible pair-wise commutations, i.e.

\begin{equation}
\label{eq:commut}
\left[X_{i}, X_{j}\right] = \lambda_{ij} X_{k}, \qquad \lambda_{ij} \in \mathbb{C}, \quad X_{k} \in \mathcal{L}
\end{equation}

The Hamiltonian of the system can be written as a finite sum of elements of the basis set with corresponding coefficients ($\gamma$) that could be constant or time-dependent. 

\begin{equation}
\label{eq:hamsum}
\mathscr{H}(t)=\sum_{i=1}^{N} \gamma_{i}(t) X_{i}
\end{equation}

Note that this is a general construction and some coefficients, $\gamma$, could be 0.

We are interested in the time evolution of individual elements of the basis set $\mathcal{L}$, the solution of which is given by the Liouville–von Neumann equation for the time evolution of the density matrix of the system, $\rho(t)$:

\begin{equation}
\label{eq:LvN}
\frac{\partial \rho(t)}{\partial t}=-\mathrm{i} [\mathscr{H}(t),\rho(t)]
\end{equation}

Similar to the Hamiltonian, we write the density matrix as a finite sum of elements of the basis set $\mathcal{L}$ and their time-dependent coefficients:

\begin{equation}
\label{eq:rhosum}
\rho(t)=\sum_{j=1}^{N} g_{j}(t) X_{j}
\end{equation}

by replacing \cref{eq:hamsum} and \cref{eq:rhosum} in \cref{eq:LvN} we have:

\begin{equation}
\begin{aligned} \sum_{j=1}^{N} \dot{g_{j}}(t) X_{j} &=-\mathrm{i} \left[\sum_{i=1}^{N} \gamma_{i} X_{i},\sum_{j=1}^{N} g_{j}(t) X_{j}\right] \\ &=-\mathrm{i} \sum_{i=1}^{N} \sum_{j=1}^{N} \gamma_{i} g_{j}(t) \left[ X_{i}, X_{j}\right] \end{aligned}
\end{equation}

This equation can be written in vector and matrix form as

\begin{equation}
\label{eq: gammaxg}
\mathcal{L} \dot{\bm{g}}(t)=-\mathrm{i} \bm{\gamma} \bm{X_{[.,.]}} \bm{g}(t)
\end{equation}

where

\begin{equation}
\bm{\gamma}=[\gamma_1,\gamma_2,\cdots,\gamma_N]
\end{equation}

\begin{equation}
\label{eq:X_commut}
\bm{X_{[.,.]}}=\left(\begin{array}{cccc}{[X_1,X_1]} & {[X_1,X_2]} & {\cdots} & {[X_1,X_N]} \\ {[X_2,X_1]} & {[X_2,X_2]} & {\cdots} & {[X_2,X_N]} \\ {\vdots} & {\vdots} & {\ddots} & {\vdots} \\ {[X_N,X_1]} & {[X_N,X_2]} & {\cdots} & {[X_N,X_N]}\end{array}\right)
\end{equation}

and

\begin{equation}
\bm{g}(t)=\left[g_1(t),g_2(t),\cdots,g_N(t)\right]^{T}
\end{equation}

with $\dot{\bm{g}}(t)$ being time derivative of $\bm{g}(t)$.

By collecting and ordering coefficients of the basis set $\mathcal{L}=\left\{X_{1}, X_{2}, \ldots, X_{N}\right\}$ we can re-write the \cref{eq: gammaxg} as

\begin{equation}
\mathcal{L} \dot{\bm{g}}(t)=-\mathrm{i} \mathcal{L} \bm{\Gamma} \bm{g}(t)
\end{equation}

and therefore 

\begin{equation}
\label{eq: Xmat}
\dot{\bm{g}}(t)=-\mathrm{i} \bm{\Gamma} \bm{g}(t)
\end{equation}

where $\Gamma$ is a matrix of the Hamiltonian coefficients, elements of which can be obtained as:

\begin{equation}
\bm{\Gamma}_{mn} = \sum_{i=1}^{N} \gamma_{i} \lambda_{in}
\end{equation}

when for each pair $X_{i}$ and $X_{n}$, $\left[X_{i},X_{n}\right]=\lambda_{in} X_{m}$.


\subsubsection{Example: single spin-$\frac{1}{2}$}

For a single spin-$\frac{1}{2}$ the basis set can be written using ladder operators of Pauli matrices ($\sigma_{\alpha}, \quad \alpha \in \{x,y,z\}$):

\begin{equation}
\label{eq:basis}
\mathcal{L}=\left\{\sigma^{-},\sqrt{2} \sigma_{z},\sigma^{+}\right\}
\end{equation}

where $\sigma^{-}= \sigma_{x}-\mathrm{i} \sigma_{y}$ and $\sigma^{+}= \sigma_{x}+\mathrm{i} \sigma_{y}$.

\begin{remark}
The unitary operator is discarded from the basis set $\mathcal{L}$ for simplicity, although its inclusion will not change the mathematical approach, neither the results, presented in this article.
\end{remark}

In the presence of a chirped pulse as in \cref{eq:suchi} with $\mathcal{RF}$ amplitude ($\omega_{1}$), bandwidth ($\Delta F$), duration ($\tau_{p}$), overall phase ($\phi_{0}$), time offset ($\delta_{t}$), and frequency offset ($\delta_{f}$), the Hamiltonian can be written as:

\begin{equation}
\mathscr{H}(t)=\beta \sigma^{-}  +\beta^{*} \sigma^{+}+\Omega  \sigma_{z}
\end{equation}

where $\Omega$ is the spin resonance offset,

\begin{equation}
\begin{aligned}
\beta=\frac{1}{2} \omega_{1} \exp &\left[-2^{n+2} \left(\frac{t-\delta_{t}}{\tau_{p}}\right)^{n} \right.\\
& \left. +\mathrm{i} \left(\phi_{0}+\frac{\pi \Delta F (t-\delta_{t})^{2}}{\tau_{p}}-2 \pi \delta_{f} (t-\delta_{t})\right)\right]
\end{aligned}
\end{equation}

and $^*$ indicates complex conjugate.

The array of Hamiltonian coefficients can be written as:

\begin{equation}
\bm{\gamma}=\left[\beta, \frac{\Omega}{\sqrt{2}},\beta^{*}\right]
\end{equation}

and according to \cref{eq:X_commut} for the basis set shown in \cref{eq:basis} we have

\begin{equation}
\begin{aligned}
\bm{X_{[.,.]}}&=\left(
\begin{array}{ccc}
{[\sigma^{-},\sigma^{-}]} & {[\sigma^{-},\sqrt{2} \sigma_{z}]} & {[\sigma^{-},\sigma^{+}]} \\
{[\sqrt{2} \sigma_{z},\sigma^{-}]} & {[\sqrt{2} \sigma_{z},\sqrt{2} \sigma_{z}]} & {[\sqrt{2} \sigma_{z},\sigma^{+}]} \\
{[\sigma^{+},\sigma^{-}]} & {[\sigma^{+},\sqrt{2} \sigma_{z}]} & {[\sigma^{+},\sigma^{+}]} \\
\end{array}
\right)\\
&=\left(
\begin{array}{ccc}
 0 & \sqrt{2}\sigma^{-} & -2 \sigma_{z} \\
 -\sqrt{2}\sigma^{-}& 0 & \sqrt{2}\sigma^{+} \\
 2\sigma_{z} & -\sqrt{2}\sigma^{+} & 0 \\
\end{array}
\right)
\end{aligned}
\end{equation}

Therefore, \cref{eq: Xmat} for this system can be written as:

\begin{equation}
\label{eq:singeq}
\left(\begin{array}{l}{\dot{g}_{1}(t)} \\ {\dot{g}_{2}(t)} \\ {\dot{g}_{3}(t)}\end{array}\right)=-\mathrm{i}\left(\begin{array}{ccc}{-\Omega} & {\sqrt{2}\beta} & {0} \\ {\sqrt{2}\beta^{*}} & {0} & {-\sqrt{2}\beta} \\ {0} & {-\sqrt{2}\beta^{*}} & {\Omega}\end{array}\right)\left(\begin{array}{l}{g_{1}(t)} \\ {g_{2}(t)} \\ {g_{3}(t)}\end{array}\right)
\end{equation}

%
\subsubsection{Example: two-spin-$\frac{1}{2}$ system with scalar coupling}
\label{subsec:ex2spin}

For a two-spin-$\frac{1}{2}$ system the basis set contains 15 orthonormal terms. Again, the choice of basis set and the order of elements in the basis set is user-defined. For the following demonstration we consider the basis set as follows:

By considering

\begin{equation}
P_{\alpha}=\sigma_{\alpha} \otimes \mathbb{1}, \quad Q_{\alpha}=\mathbb{1} \otimes \sigma_{\alpha}, \qquad \alpha \in \{x,y,z\}
\end{equation}

and 

\begin{equation}
P^{\pm}=P_{x} \pm \mathrm{i} P_{y}, \qquad Q^{\pm}=Q_{x} \pm \mathrm{i} Q_{y}
\end{equation}

we can write a basis set consisting of all 15 terms, including 5 zero quantum, 8 single quantum, and 2 double quantum terms.

\begin{equation}
\label{eq:basis2}
\begin{aligned}
\mathcal{L}= & \left\{  P_z, Q_z, P^{-} Q^{+}, P^{+} Q^{-}, 2 P_{z} Q_{z}, \right.\\
& \frac{P^{-}}{\sqrt{2}}, \frac{P^{+}}{\sqrt{2}}, \frac{Q^{-}}{\sqrt{2}}, \frac{Q^{+}}{\sqrt{2}}, \sqrt{2}  P^{-} Q_{z}, \sqrt{2} P^{+}Q_{z}, \sqrt{2} P_{z}Q^{-}, \sqrt{2} P_{z}Q^{+}, \\
& \left. P^{-} Q^{-}, P^{+}Q^{+} \right\}
\end{aligned}
\end{equation}

The Hamiltonian for this system under a chirped pulse with a set of parameters specified in the previous section can be written as:

\begin{equation}
\label{eq:ham2}
\begin{aligned}
\mathscr{H}(t) = & \Omega_{P} P_{z} + \Omega_{Q} Q_{z} + \pi J \left(  P^{-} Q^{+} +  P^{+} Q^{-} + 2 P_{z} Q_{z} \right) \\
& + \beta \left(  P^{-} +  Q^{-} \right)  + \beta^{*} \left(  P^{+} +  Q^{+} \right)
\end{aligned}
\end{equation}

According to \cref{eq:basis2} and \cref{eq:ham2} the coefficient array, $\gamma$, can be written as:

\begin{equation}
\label{eq:coeff2}
\begin{aligned}
\bm{\gamma}= & \left[ \Omega_{P},\Omega_{Q}, \pi J, \pi J, \pi J, \right.\\
& \sqrt{2} \beta, \sqrt{2} \beta^{*}, \sqrt{2} \beta, \sqrt{2} \beta^{*}, 0, 0, 0, 0, \\
& \left. 0, 0 \right]
\end{aligned}
\end{equation}

After constructing $\bm{X_{[.,.]}}$ (\cref{eq:X_commut}) using all pairwise commutations of the basis set $\mathcal{L}$ for this system, the homogeneous ODE of \cref{eq: Xmat} for the time coefficients of the basis set can be written as \cref{eq:ode2col}.

where 

\begin{equation*}
\mathcal{J}=\pi J, \quad \mathcal{B} = \sqrt{2}\beta, \quad \Sigma = \Omega_{P} + \Omega_{Q}, \quad \Delta = \Omega_{P} - \Omega_{Q}
\end{equation*}

\begin{widetext}
\begin{equation}
\label{eq:ode2col}
\mathrm{i} \left(\begin{array}{l}{\dot{g}_{1}(t)} \\ {\dot{g}_{2}(t)} \\ {\dot{g}_{3}(t)} \\ {\dot{g}_{4}(t)} \\ {\dot{g}_{5}(t)} \\ {\dot{g}_{6}(t)}\\ {\dot{g}_{7}(t)} \\ {\dot{g}_{8}(t)} \\ {\dot{g}_{9}(t)}\\ {\dot{g}_{10}(t)} \\ {\dot{g}_{11}(t)} \\ {\dot{g}_{12}(t)}\\ {\dot{g}_{13}(t)} \\ {\dot{g}_{14}(t)} \\ {\dot{g}_{15}(t)}\end{array}\right)=\left(
\begin{array}{ccccccccccccccc}
 0 & 0 & \mathcal{J} & -\mathcal{J} & 0 & \mathcal{B} & -\mathcal{B}^{*} & 0 & 0 & 0 & 0 & 0 & 0 & 0 & 0 \\
 0 & 0 & -\mathcal{J} & \mathcal{J} & 0 & 0 & 0 & \mathcal{B} & -\mathcal{B}^{*} & 0 & 0 & 0 & 0 & 0 & 0 \\
 \mathcal{J} & -\mathcal{J} & -\Delta & 0 & 0 & 0 & 0 & 0 & 0 & -\mathcal{B} & 0 & 0 & \mathcal{B}^{*} & 0 & 0 \\
-\mathcal{J} & \mathcal{J} & 0 & \Delta & 0 & 0 & 0 & 0 & 0 & 0 & \mathcal{B}^{*} & -\mathcal{B} & 0 & 0 & 0 \\
 0 & 0 & 0 & 0 & 0 & 0 & 0 & 0 & 0 & \mathcal{B} & -\mathcal{B}^{*} & \mathcal{B} & -\mathcal{B}^{*} & 0 & 0 \\
 \mathcal{B}^{*} & 0 & 0 & 0 & 0 & -\Omega _1 & 0 & 0 & 0 & -\mathcal{J} & 0 & \mathcal{J} & 0 & 0 & 0 \\
 -\mathcal{B} & 0 & 0 & 0 & 0 & 0 & \Omega _1 & 0 & 0 & 0 & \mathcal{J} & 0 & -\mathcal{J} & 0 & 0 \\
 0 & \mathcal{B}^{*} & 0 & 0 & 0 & 0 & 0 & -\Omega _2 & 0 & \mathcal{J} & 0 & -\mathcal{J} & 0 & 0 & 0 \\
 0 & -\mathcal{B} & 0 & 0 & 0 & 0 & 0 & 0 & \Omega _2 & 0 & -\mathcal{J} & 0 & \mathcal{J} & 0 & 0 \\
 0 & 0 & -\mathcal{B}^{*} & 0 & \mathcal{B}^{*} & -\mathcal{J} & 0 & \mathcal{J} & 0 & -\Omega _1 & 0 & 0 & 0 & \mathcal{B} & 0 \\
 0 & 0 & 0 & \mathcal{B} & -\mathcal{B} & 0 & \mathcal{J} & 0 & -\mathcal{J} & 0 & \Omega _1 & 0 & 0 & 0 & -\mathcal{B}^{*} \\
 0 & 0 & 0 & -\mathcal{B}^{*} & \mathcal{B}^{*} & \mathcal{J} & 0 & -\mathcal{J} & 0 & 0 & 0 & -\Omega _2 & 0 & \mathcal{B} & 0 \\
 0 & 0 & \mathcal{B} & 0 & -\mathcal{B} & 0 & -\mathcal{J} & 0 & \mathcal{J} & 0 & 0 & 0 & \Omega _2 & 0 & -\mathcal{B}^{*} \\
 0 & 0 & 0 & 0 & 0 & 0 & 0 & 0 & 0 & \mathcal{B}^{*} & 0 & \mathcal{B}^{*} & 0 & -\Sigma & 0 \\
 0 & 0 & 0 & 0 & 0 & 0 & 0 & 0 & 0 & 0 & -\mathcal{B} & 0 & -\mathcal{B} & 0 & \Sigma \\
\end{array}
\right) \left(\begin{array}{l}{g_{1}(t)} \\ {g_{2}(t)} \\ {g_{3}(t)} \\ {g_{4}(t)} \\ {g_{5}(t)} \\ {g_{6}(t)}\\ {g_{7}(t)} \\ {g_{8}(t)} \\ {g_{9}(t)}\\ {g_{10}(t)} \\ {g_{11}(t)} \\ {g_{12}(t)}\\ {g_{13}(t)} \\ {g_{14}(t)} \\ {g_{15}(t)}\end{array}\right)
\end{equation}
\end{widetext}


\subsubsection{Demo: zero-quantum suppression}

The elegant idea of the suppression of zero quantum coherences using spatiotemporal averaging was first introduced by Thrippleton and Keeler  \cite{Thrippleton2003}. Here the general attenuation mechanism of this method on a two-spin-$\frac{1}{2}$ system is presented. The pulse element simply consists of a $180^{\circ}$ chirped pulse applied simultaneously with a pulsed field gradient. The maximum $\mathcal{RF}$ amplitude of a chirped pulse can be calculated as:

\begin{equation}
\label{eq:rfqchirp}
\mathcal{RF}_{max}=\sqrt{\frac{\Delta F \mathcal{Q} }{2 \pi \tau_{p}}}
\end{equation}

Where $\mathcal{Q}$ is the adiabaticity factor. The relationship between $\mathcal{Q}$ factor and pulse flip angle $\alpha$ can be written as \cite{Jeschke2015}:

\begin{equation}
\mathcal{Q}=\frac{2}{\pi}\ln{\left(\frac{2}{\cos{(\alpha)}+1}\right)}
\end{equation}

The effective flip angle of a chirped pulse approaches $180^{\circ}$ asymptotically as the $\mathcal{RF}$ increases, therefore a value of $\mathcal{Q}$ factor (5 for most practical purposes) is chosen as a threshold in order to satisfy the adiabatic condition with an affordable $\mathcal{RF}$ amplitude.

The Hamiltonian for two coupled spins in the presence of a chirped pulse and simultaneous pulsed field gradient can be written as:

\begin{equation}
\begin{aligned}
\mathscr{H}(t)=&\Omega_{P} P_{z}+\Omega_{Q} Q_{z} \\
&+\pi J\left(P^{-} Q^{+}+P^{+} Q^{-}+2 P_{z} Q_{z}\right) \\
&+\beta \left(  P^{-} +  Q^{-} \right)  + \beta^{*} \left(  P^{+} +  Q^{+} \right)\\
&+\Omega_{g}(z) \left(P_{z} +Q_{z}\right)
\end{aligned}
\end{equation}

Since for a single chirped pulse here $\delta_{t}=0$, $\delta_{f}=0$, and $\phi_{0}=0$:  

\begin{equation}
\beta = \frac{1}{2} \omega_{1} \exp \left[-2^{n+2} \left(\frac{t-\frac{\tau_{p}}{2}}{\tau_{p}}\right)^{n} +\mathrm{i} \left(\frac{\pi \Delta F \left(t-\frac{\tau_{p}}{2}\right)^{2}}{\tau_{p}}\right)\right] 
\end{equation}

The bandwidth of the frequency distribution induced by the field gradient is 

\begin{equation}
\Delta \Omega_{g} = \gamma_{\text{H}} L G_{z}, 
\end{equation}

where $\gamma_{\text{H}}$ is the gyromagnetic ratio of proton and $L$ is the length of the active volume exposed to a pulsed field gradient along $z$ axis ($G_{z}$). The frequency offset induced at each part of the active volume can be written as 

\begin{equation}
\label{eq:swg}
\Omega_{g}(z)=\gamma_{\text{H}} z G_{z}, \quad z \in \left[-\frac{L}{2},\frac{L}{2}\right].
\end{equation}

For practical implementation in order to avoid unnecessary signal loss due to diffusion or excessive spatial encoding normally the strength of the gradient $G_{z}$ is set to a value so that $\Omega_{g}(z) \in \left[-\pi \Delta F, +\pi \Delta F\right]$.

The array of coefficients for this system can be written as:

\begin{equation}
\label{eq:coeff2chzq}
\begin{aligned}
\bm{\gamma}= & \left[ \Omega_{P}+\Omega_{g}(z),\Omega_{Q}+\Omega_{g}(z), \pi J, \pi J, \pi J, \right.\\
& \sqrt{2} \beta, \sqrt{2} \beta^{*}, \sqrt{2} \beta, \sqrt{2} \beta^{*}, 0, 0, 0, 0, \\
& \left. 0, 0 \right]
\end{aligned}
\end{equation}

Now including these in \cref{eq:ode2col} we can set the ODE for this system. We need to replace $\Omega_{P}$ and $\Omega_{Q}$ with $\Omega_{P} + \Omega_{g}(z)$ and $\Omega_{Q} + \Omega_{g}(z)$ respectively. Additionally we have:

\begin{equation*}
\mathcal{J}=\pi J, \quad \mathcal{B} =\sqrt{2}\beta, \quad \Sigma = \Omega_{P} + \Omega_{Q} + 2\Omega_{g}(z), \quad \Delta = \Omega_{P} - \Omega_{Q}
\end{equation*}

In order to gain useful insight about the mechanism of zero quantum suppression using spatiotemporal averaging, let us define a parameter $\xi$ indicating a unique relationship between spin system and pulse element parameters:

\begin{equation}
\label{eq:xisweep}
\xi=\sqrt{\frac{\theta}{\mathcal{R}}}
\end{equation}

where 

\begin{equation*}
\theta=\left|\frac{\Delta}{2 \pi J}\right|, \quad \mathcal{R}=\frac{\Delta F}{\tau_{p}}
\end{equation*}

Note that here $\mathcal{R}$ is the sweep-rate of the pulse and $\theta$ indicates the strength of the interaction between spins. As the phase of a zero quantum coherence is time dependent, a combination of linear-sweep irradiation of spins with pulsed field gradient induces a time-dependent phase variation of unwanted zero quantum coherences across the sample volume. In order to demonstrate the effect of this pulse element, let us consider that only one zero quantum term $P^{-}Q^{+}$ exists before the pulse element, i.e. $g_{3}(0)=1$. As the chirped pulse is $180^{\circ}$ we are interested to see how much $g_{4}(\tau_{p})$ corresponding to term $P^{+}Q^{-}$ is generated at the end of the pulse element.

\begin{figure}
  \includegraphics[width=\linewidth]{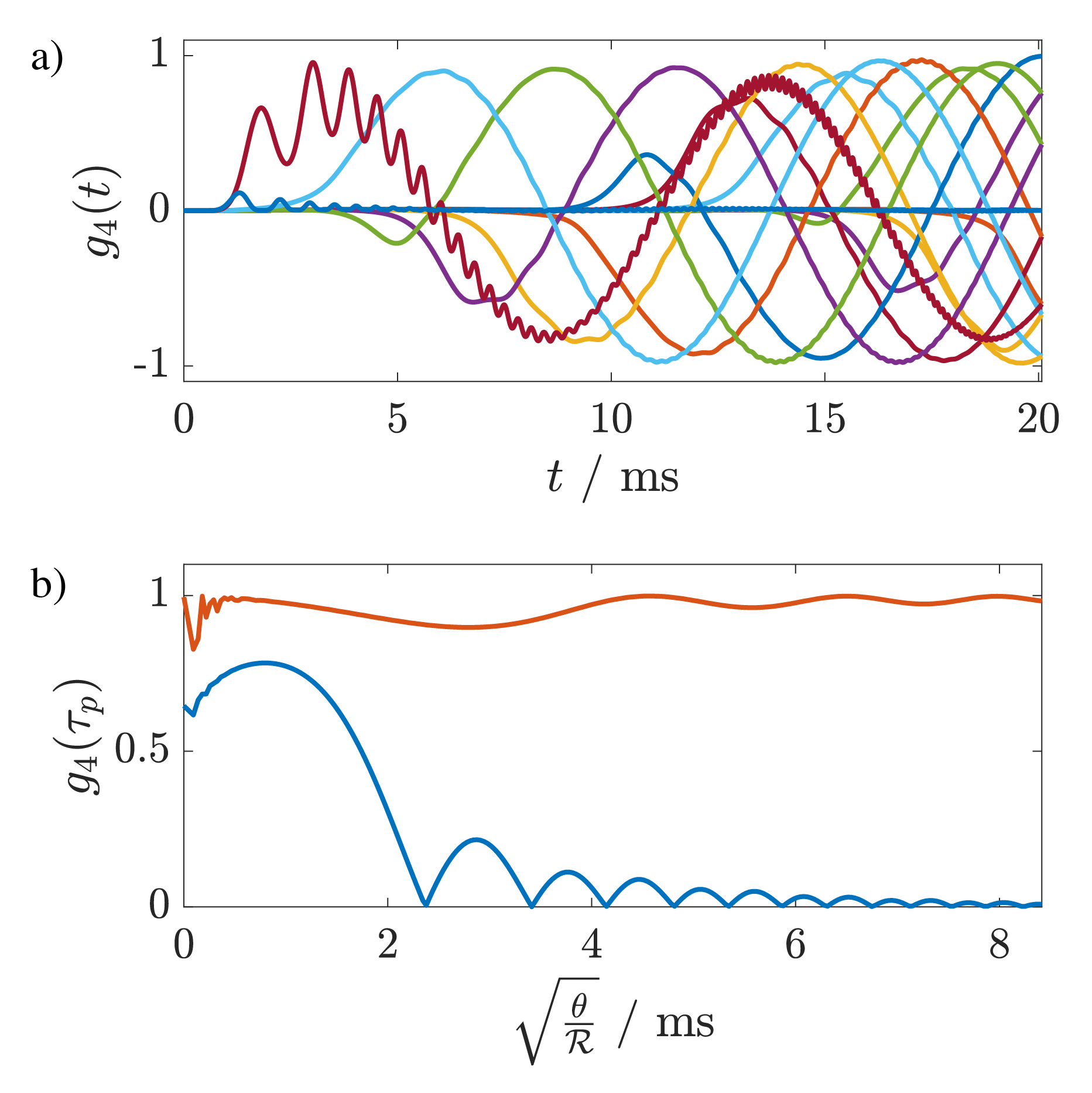}\\
  \caption{Graphical representation of zero quantum coherence during a chirped pulse applied simultaneously with a pulsed field gradient; (a) time variation of $g_{4}(t)$ corresponding to zero quantum term $P^{+}Q^{-}$ for 15 different values of $\Omega_{g}(z)$ during a pulse element with $\tau_{p}=20$ ms, $\Delta F = 10000$ Hz, $\Omega_{g} = 2 \pi 10000$, (b) time variation of $g_{4}(\tau_{p})$ as a function of $\xi$ in the absence (red) and presence (blue) of the pulsed field gradient.}
  \label{fig:fig_3}
\end{figure}

\Cref{fig:fig_3}(a) shows the variation of $g_{4}(t)$ during a zero quantum filter at different parts of the sample volume. It is evident that zero quantum signals at different parts of the sample acquire time-dependent phase shift leading of an efficient attenuation of those signals via ensemble averaging. \Cref{fig:fig_3}(b) represents the efficiency of the zero quantum attenuation by plotting $g_{4}(\tau_{p})$ as a function of $\xi$ in \cref{eq:xisweep} in the presence and absence of spatial averaging. The blue graph in \cref{fig:fig_3}(b) suggests that for example for a two-spin system with 200 Hz difference in offsets and 10 Hz $J$ coupling, in order to attenuate the zero-quantum term to 3\% with a 10 kHz chirped pulse, the duration of the pulse should be 18 ms.


\subsection{Wei-Norman Lie algebra} 
\label{subsec:wnla}

One elegant way of solving problems presented here was proposed by Wei and Norman \cite{Wei1963,Wei1964}. Here the goal is to obtain a solution for a time-ordered propagator $\mathscr{U}(t)$ rather than the density matrix. The solution of the Liouville–von Neumann equation can be written using a time-dependent unitary propagator as:

\begin{equation}
\label{eq:urhou}
\rho(t)= \mathscr{U}(t) \rho(0) \mathscr{U}(t)^{\dagger}
\end{equation}

Where $\mathscr{U}(t)$ is the solution of the following differential equation: 

\begin{equation}
\label{eq:uode}
\frac{d}{d t} \mathscr{U}(t)=-\mathrm{i} \mathscr{H}(t) \mathscr{U}(t), \quad \mathscr{U}(0)=\mathbb{1}
\end{equation}

with the solution

\begin{equation}
\mathscr{U}(t)=\mathscr{T} \exp \left[-\mathrm{i}  \int_{0}^{t} \mathscr{H}\left(t^{\prime}\right) d t^{\prime}\right]
\end{equation}

where $\mathscr{T}$ is the time-ordering operator. The goal is, relying on the Lie algebraic properties of the orthonormal basis set and the fact that the set is closed under all possible commutations, to write the solution of this differential equation as product of the elements of the Lie set with some time-dependent coefficients.

Considering a basis set 

\begin{equation}
\mathcal{L}=\left\{X_{1}, X_{2}, \ldots, X_{N}\right\}
\end{equation}

the elements of which satisfy the condition of \cref{eq:commut}, the assumed solution of the time-dependent propagator $\mathscr{U}(t)$ in \cref{eq:uode} can be written as:

\begin{equation}
\label{eq:uprod}
\mathscr{U}(t)=\prod_{i=1}^{N} \exp \left[g_{i}(t) X_{i}\right] ; \quad g_{i}(0)=0
\end{equation}

where $g_{i}(t)$ is a smooth function of time that gives the time variation of $X_{i}$ element of the basis set $\mathcal{L}$.

By replacing \cref{eq:hamsum} and \cref{eq:uprod} in \cref{eq:uode} and differentiation we obtain

\begin{equation}
\label{eq:gdotinu}
\sum_{k=1}^{N} \dot{g}_{k}(t) \mathcal{E}_{1}^{k-1} X_{k} \mathcal{E}_{k}^{N} =-\mathrm{i}  \left(\sum_{j=1}^{N} \gamma_{j} X_{j}\right) \mathcal{E}_{1}^{N}
\end{equation}

where

\begin{equation}
\label{eq:expprod}
    \mathcal{E}_{p}^{q} =  
\begin{dcases}
     \prod_{i=p}^{q} \exp \left(g_{i}(t) X_{i}\right),& \text{if } q > p\\
     \prod_{i=p}^{q} \exp \left(-g_{i}(t) X_{i}\right),& \text{if } q< p
\end{dcases}
\end{equation}

Note that the products of exponentials in \cref{eq:expprod} are time-ordered, the order of which is dictated by the initial choice of the basis set $\mathcal{L}$.

By multiplying both sides of \cref{eq:gdotinu} by $\mathcal{E}_{N}^{1}$ we obtain

\begin{equation}
\sum_{k=1}^{N} \dot{g}_{k}(t) \mathcal{E}_{1}^{k-1} X_{k} \mathcal{E}_{k-1}^{1} =-\mathrm{i}  \sum_{j=1}^{N} \gamma_{j} X_{j}
\end{equation}

We can write this equation in vector and matrix form as

\begin{equation}
\mathcal{L} \Xi(g) \dot{\bm{g}}(t) =- \mathrm{i} \mathcal{L} \bm{\gamma}^{T}
\end{equation}

and therefore 

\begin{equation}
\label{eq:no_inv}
\Xi(g) \dot{\bm{g}}(t) =- \mathrm{i} \bm{\gamma}^{T}
\end{equation}

Here the $k^{th}$ column of the matrix $\Xi(g)$ will be ordered coefficients of $\mathcal{L}$ in $\mathcal{E}_{1}^{k-1} X_{k} \mathcal{E}_{k-1}^{1}$. Construction of the matrix $\Xi$ will be discussed in details in \cref{subsec:exwnla}.

It is known from BCH (Baker–Campbell–Hausdorff) formula that the expression $\mathcal{E}_{1}^{k-1} X_{k} \mathcal{E}_{k-1}^{1}$ can be written using a series consisting of nested commutators:

\begin{equation}
\label{eq:bch}
e^{X} Y e^{-X}=Y + [X,Y] +  \frac{[X,[X,Y]]}{2!} + \frac{[X,[X,[X,Y]]]}{3!}+\cdots
\end{equation}

A linear adjoint operator and its powers can be defined as:

\begin{equation}
\mathrm{ad}_X^0 (Y) = X, \quad \mathrm{ad}_X(Y) = [X,Y], \quad \mathrm{ad}_X^2(Y) = [X,[X,Y]], \cdots
\end{equation}

Therefore \cref{eq:bch} can be written in terms of adjoint exponentials:

\begin{equation}
e^{X} Y e^{-X}=(e^{\mathrm{ad}_X}) Y
\end{equation}

As it will be shown in \cref{subsec:exwnla}, the series in \cref{eq:bch} can be written as a closed-form generating function. For any two members, $X_{i}$ and $X_{j}$ of a basis set $\mathcal{L}$, that satisfy condition in \cref{eq:commut} we have:

\begin{equation}
\label{eq:genfun}
   e^{g X_{j}} X_{i} e^{- g X_{j}} =  
\begin{dcases}
     X_{i},& \text{if } [X_{i},X_{j}]=0\\
     X_{i} \cosh(g) + [X_{j},X_{i}] \sinh(g) ,& \text{if } [X_{i},X_{j}]\neq0
\end{dcases}
\end{equation}

It can be shown that the matrix $\Xi$ which is an analytic function of time-dependent coefficients $g_{i}(t)$ is always invertible around time zero given the fact that at $t=0$ $\Xi$ is the Identity matrix and there is always a neighbourhood where it has non-zero determinant. Therefore, \cref{eq:no_inv} can be re-arranged and be written as a system of differential equations as:

\begin{equation}
\label{eq:eqwn}
\dot{\bm{g}}(t) =- \mathrm{i} {\Xi^{-1}(g)}\bm{\gamma}^{T}
\end{equation}


\subsubsection{Example: single spin-$\frac{1}{2}$}
\label{subsec:exwnla}

In this section, a comprehensive example for the construction of \cref{eq:eqwn} for a single spin-$\frac{1}{2}$ under a chirped pulse will be presented. Let us consider the Cartesian basis set as:

\begin{equation}
\mathcal{L}=\left\{\sigma_{x}, \sigma_{y}, \sigma_{z}\right\}
\end{equation}

Although the order of the elements in the basis set can be arbitrarily chosen, we need to use the same order throughout our calculations.

We can write a set of cyclic commutations for a single spin-$\frac{1}{2}$ as follows:

\begin{equation}
[\sigma_{x},\sigma_{y}] = \mathrm{i} \sigma_{z}, \quad [\sigma_{y},\sigma_{z}] = \mathrm{i} \sigma_{x}, \quad [\sigma_{z},\sigma_{x}] = \mathrm{i} \sigma_{y} 
\end{equation}

the Hamiltonian of the system under a chirped pulse can be written as:

\begin{equation}
\label{eq:hamwn}
\mathscr{H}(t)=\Omega  \sigma_{z} + C_{x}(t) \sigma_{x} + C_{y}(t) \sigma_{y}
\end{equation}

and 

\begin{equation}
\label{eq:coefofham}
\bm{\gamma}=\left[C_{x}(t), C_{y}(t), \Omega\right].
\end{equation}

where $C_{x}(t)$ and $C_{y}(t)$ are real and imaginary parts of the chirped pulse function in \cref{eq:suchi} respectively. This will be discussed in great details in \cref{subsec:democomp} and \cref{subsec:chorus}.

We consider the solution to \cref{eq:uode} as follows:

\begin{equation}
\label{eq:usol}
\mathscr{U}(t)=e^{g_{1}(t)\sigma_{x}} e^{g_{2}(t)\sigma_{y}} e^{g_{3}(t)\sigma_{z}} 
\end{equation}

where $g_{1}(t)$, $g_{2}(t)$, and $g_{3}(t)$  are solutions of \cref{eq:eqwn}. This approach allows us to write the total propagator of the pulse sequence as an ordered product of time-dependent orthogonal rotation matrices, which is an attractive feature for the study and control of spin systems under field modulations \cite{Campolieti1989} and is compatible with rotation-operator disentanglement approach \cite{Zhou1994}.

The next step would be constructing the matrix $\Xi(g_{1},g_{2},\cdots,g_{N})$. According to \cref{eq:no_inv} and the order of elements in the basis set $\mathcal{L}$, columns of $\Xi(g_{1},g_{2},\cdots,g_{N})$ can be obtained as functions of coefficients of corresponding elements of the basis.

Therefore three columns of $\Xi(g_{1},g_{2},\cdots,g_{N})$ will be obtained from:

\begin{equation}
\begin{dcases}
     \text{1}^\text{st}: \qquad \sigma_{x}\\
     \text{2}^\text{nd}: \qquad e^{g_{1}(t)\sigma_{x}}\sigma_{y} e^{-g_{1}(t)\sigma_{x}}\\
     \text{3}^\text{rd}: \qquad e^{g_{1}(t)\sigma_{x}} e^{g_{2}(t)\sigma_{y}}  \sigma_{z} e^{-g_{2}(t)\sigma_{y}} e^{-g_{1}(t)\sigma_{x}} \\
\end{dcases}
\end{equation}

It is clear that the $\text{1}^\text{st}$ column is trivial. The $\text{2}^\text{nd}$ column can be obtained via the application of \cref{eq:bch} and the $\text{3}^\text{rd}$ column is the nested version of \cref{eq:bch}. Note that this is a general approach and can be applied to any basis set that obeys the Lie algebra of $\mathrm{SU}(2^{N})$ groups.

As an example, we derive the second column explicitly using \cref{eq:bch} and show how the solution can be represented as a closed-form generating function presented in \cref{eq:genfun}. The first four nested commutators used in \cref{eq:bch} can be written as:

\begin{equation}
\label{eq:c2nestcumm}
\begin{dcases}
     [g_{1}(t)\sigma_{x},\sigma_{y}]=\mathrm{i} \sigma_{z} g_{1}(t)\\
     [g_{1}(t)\sigma_{x},[g_{1}(t)\sigma_{x},\sigma_{y}]]=\mathrm{i} \sigma_{y} g_{1}^{2}(t)\\
     [g_{1}(t)\sigma_{x},[g_{1}(t)\sigma_{x},[g_{1}(t)\sigma_{x},\sigma_{y}]]]=\mathrm{i} \sigma_{z} g_{1}^{3}(t)\\
     [g_{1}(t)\sigma_{x},[g_{1}(t)\sigma_{x},[g_{1}(t)\sigma_{x},[g_{1}(t)\sigma_{x},\sigma_{y}]]]]=\mathrm{i} \sigma_{y} g_{1}^{4}(t)
\end{dcases}
\end{equation}

therefore \cref{eq:bch} can be written as a series and its closed-form generation function involving hyperbolic sines and cosines.

\begin{equation}
\label{eq:c2bch}
\begin{aligned}
(e^{\mathrm{ad}_{g_{1}(t)\sigma_{x}}}) \sigma_{y}&=\sigma_{y} + \mathrm{i} \sigma_{z} g_{1}(t) +  \frac{\mathrm{i} \sigma_{y} g_{1}^{2}(t)}{2!} + \frac{\mathrm{i} \sigma_{z} g_{1}^{3}(t)}{3!}+\cdots\\
&=\sigma_{y} \cosh(g_{1}(t)) + \mathrm{i} \sigma_{z} \sinh(g_{1}(t))
\end{aligned}
\end{equation}

\Cref{eq:c2bch} can be written in a general form as \cref{eq:genfun}.

Similarly, we can obtain elements of the last column of $\Xi(g_{1},g_{2},\cdots,g_{N})$:

\begin{equation}
\begin{aligned}
(e^{\mathrm{ad}_{g_{1}(t)\sigma_{x}}}) (e^{\mathrm{ad}_{g_{2}(t)\sigma_{y}}}) \sigma_{z} &= \mathrm{i} \sigma_{x} \sinh(g_{2}(t))\\
&-\mathrm{i}\sigma_{y}\sinh(g_{1}(t))\cosh(g_{2}(t))\\
&+\sigma_{z}\cosh(g_{1}(t))\cosh(g_{2}(t))
\end{aligned}
\end{equation}

Finally the matrix $\Xi(g)$ is obtained as:

\begin{equation}
\label{eq:xibig}
\Xi=\left(\begin{array}{ccc}{1} & {0} & {\mathrm{i}\sinh(g_{2}(t))} \\ {0} & {\cosh(g_{1}(t))} & {-\mathrm{i}\sinh(g_{1}(t))\cosh(g_{2}(t))} \\ {0} & {\mathrm{i}\sinh(g_{1}(t))} & {\cosh(g_{1}(t))\cosh(g_{2}(t))}\end{array}\right)
\end{equation}

and the inverse of $\Xi$ required for \cref{eq:eqwn} is:

\begin{equation*}
\left(\begin{array}{ccc}{1} & {-\sinh(g_{1}(t))\tanh(g_{2}(t))} & {-\mathrm{i}\cosh(g_{1}(t))\tanh(g_{2}(t))} \\ {0} & {\cosh(g_{1}(t))} & {\mathrm{i}\sinh(g_{1}(t))} \\ {0} & {-\mathrm{i}\sinh(g_{1}(t)) \sech(g_{2}(t))} & {\cosh(g_{1}(t)) \sech(g_{2}(t))}\end{array}\right)
\end{equation*}

Finally, using \cref{eq:coefofham,eq:xibig} we can write \cref{eq:eqwn} as:

\begin{equation}
\label{eq:wnode}
\left(\begin{array}{l}{\dot{g}_{1}} \\ {\dot{g}_{2}} \\ {\dot{g}_{3}}\end{array}\right)=\left(
\begin{array}{c}
-\mathrm{i} C_{x}+ \tanh \left(g_2\right) \left(-\Omega  \cosh \left(g_1\right)+\mathrm{i} C_{y} \sinh \left(g_1\right)\right) \\
 \Omega  \sinh \left(g_1\right)-\mathrm{i} C_{y} \cosh \left(g_1\right) \\
 -\sech\left(g_2\right) \left(C_{y} \sinh \left(g_1\right)+\mathrm{i} \Omega  \cosh \left(g_1\right)\right) \\
\end{array}
\right)
\end{equation}

Which can be solved to obtain time-dependent coefficients, and hence the total time-dependent propagator $\mathscr{U}(t)$.

Of course it would be more desirable to have access to the structure of the density matrix as a result of such propagator. This can be achieved straightforwardly by the direct application of \cref{eq:usol} in \cref{eq:urhou}.

\begin{equation}
\label{eq:rhown}
\rho(t)=e^{g_{1}(t)\sigma_{x}} e^{g_{2}(t)\sigma_{y}} e^{g_{3}(t)\sigma_{z}}\rho(0) e^{-g_{3}(t)\sigma_{z}} e^{-g_{2}(t)\sigma_{y}} e^{-g_{1}(t)\sigma_{x}}
\end{equation}

where

\begin{equation}
\label{eq:rhoc}
\begin{aligned}
&\rho(t)=c_{1}(t) \sigma_{x} + c_{2}(t) \sigma_{y} +c_{3}(t) \sigma_{z},\\ 
&\rho(0)=c_{1}(0) \sigma_{x} + c_{2}(0) \sigma_{y} +c_{3}(0) \sigma_{z}
\end{aligned}
\end{equation}

Similar to previous sections, we can write \cref{eq:rhown} in vector and matrix form:

\begin{equation}
\label{eq:rhogamma}
\left(\begin{array}{l}{c_{1}(t)} \\ {c_{2}(t)} \\ {c_{3}(t)}\end{array}\right)= \bm{\Gamma}(t) \left(\begin{array}{l}{c_{1}(0)} \\ {c_{2}(0)} \\ {c_{3}(0)} \end{array}\right)
\end{equation}

Here rows of the matrix $\bm{\Gamma}$ is obtained using the application of \cref{eq:genfun} in \cref{eq:rhown}. The general form of $\bm{\Gamma}$ in \cref{eq:rhogamma} can be obtained explicitly as \cref{eq:gamma2col}. Here the time-dependency of $g_{1}$, $g_{2}$, and $g_{3}$ is dropped for simplicity. Using \cref{eq:rhogamma} and \cref{eq:gamma2col} the state of the system at any time $\tau$ can be obtained by multiplying $\bm{\Gamma}(\tau)$ by the initial state of the system. 

\begin{figure*}
\begin{equation}
\label{eq:gamma2col}
\bm{\Gamma}=\left(
\begin{array}{ccc}{\cosh(g_{2})\cosh(g_{3})} & {-\mathrm{i}\cosh(g_{2})\sinh(g_{3})} & {\mathrm{i}\sinh(g_{2})} \\ 
{-\sinh(g_{1})\sinh(g_{2})\cosh(g_{3})+\mathrm{i}\cosh(g_{1})\sinh(g_{3})} & {\cosh(g_{1})\cosh(g_{3})+\mathrm{i}\sinh(g_{1})\sinh(g_{2})\sinh(g_{3})} & {-\mathrm{i}\sinh(g_{1})\cosh(g_{2})} \\ 
{-\mathrm{i}\cosh(g_{1})\sinh(g_{2})\cosh(g_{3})-\sinh(g_{1})\sinh(g_{3})} & {\mathrm{i}\sinh(g_{1})\cosh(g_{3})-\cosh(g_{1})\sinh(g_{2})\sinh(g_{3})} & {\cosh(g_{1}) \cosh(g_{2})}\end{array}
\right)
\end{equation}
\end{figure*}


\subsubsection{Demo: composite chirps for refocusing}
\label{subsec:democomp}

In this section, the solution of spin dynamics for a sequence of chirped pulses will be presented. This sequence, introduced by Hwang et al. \cite{Hwang1997}, offers an ideal refocusing of spins without any phase rolls across the spectral range of interest. It takes advantage of amplitude- and time-matching of three $180^{\circ}$ chirped pulses in order to achieve zero net chemical shift evolution during the sequence and self-compensation with respect to variations of the $B_{1}$ field. Here the goal is to demonstrate how one can use the proposed approach in the previous section to compute and visualise the spin dynamics during such a pulse sequence.

For this example we consider a 1 ms - 2 ms - 1 ms sequence of chirped pulses with 200 kHz bandwidth. The whole sequence can be written as a sum of three chirped waveforms using \cref{eq:genchirp,eq:genchirpx,eq:genchirpy}. Note that for this application all three pulses have the same bandwidth ($\Delta F$) and $\mathcal{RF}$ amplitudes ($\omega_{1}$), and for all of them $\phi_{0}=0$ and $\delta_{f} = 0$, but durations, $\tau_{p}$ and positions $\delta_{t}$ are different.

The total pulse sequence can therefore be written as:

\begin{equation}
\label{ctcomp}
S(t)=\omega_{1} \sum_{i=1}^{3} \exp \left[-2^{n+2} \left(\frac{t-\delta_{t}^{(i)}}{\tau_{p}^{(i)}}\right)^{n}+\mathrm{i} \left(\frac{\pi \Delta F (t-\delta_{t}^{(i)})^{2}}{\tau_{p}^{(i)}}\right)\right]
\end{equation}

The Hamiltonian of a single spin-$\frac{1}{2}$ under this sequence can be written as:

\begin{equation}
\label{eq:hamcomp}
\mathscr{H}(t)=\Omega  \sigma_{z} +C_{x}(t) \sigma_{x}+C_{y}(t) \sigma_{y}
\end{equation}

where

\begin{equation}
\label{eq:cxcycomp}
\begin{aligned}
&C_{x}(t)=\omega_{1} \sum_{i=1}^{3} \exp \left[-2^{n+2} \left(\frac{t-\delta_{t}^{(i)}}{\tau_{p}^{(i)}}\right)^{n}\right] \cos\left[\left(\frac{\pi \Delta F (t-\delta_{t}^{(i)})^{2}}{\tau_{p}^{(i)}}\right)\right]\\
&C_{y}(t)=\omega_{1} \sum_{i=1}^{3} \exp \left[-2^{n+2} \left(\frac{t-\delta_{t}^{(i)}}{\tau_{p}^{(i)}}\right)^{n}\right] \sin\left[\left(\frac{\pi \Delta F (t-\delta_{t}^{(i)})^{2}}{\tau_{p}^{(i)}}\right)\right]
\end{aligned}
\end{equation}

Now we can write all 18 parameters of the sequence (3 pulses, 6 parameters each). $\omega_{1} = 2 \pi \mathcal{RF}$ can be calculated using \cref{eq:rfqchirp}, where $\Delta F = 200000$ Hz and $\mathcal{Q} = 5$. $\delta_{t}$ is time offset of the centre-point of each pulse from the beginning of the sequence ($t=0$). As none of the three pulses here have additional phase or frequency offset, these parameters will be set to zero. Therefore, a complete set of parameters for three chirped pulses in this sequence will be as follows:

\begin{equation}
\label{eq:compchirpparam}
\begin{dcases}
     \bm{\omega}_{1} = [79267, 79267, 79267] \quad (\nicefrac{\text{rad}}{\text{s}})\\
     \bm{\Delta F} = [200000, 200000, 200000] \quad (\text{Hz})\\
     \bm{\tau}_{p} = [0.001, 0.002, 0.001] \quad (\text{s})\\
     \bm{\phi}_{0} = [0, 0, 0] \quad (\text{rad})\\
     \bm{\delta}_{t} = [0.0005, 0.0020, 0.0035] \quad (\text{s})\\
     \bm{\delta}_{f}  = [0, 0, 0] \quad (\text{Hz})\\
\end{dcases}
\end{equation}

\begin{figure}
  \includegraphics[width=\linewidth]{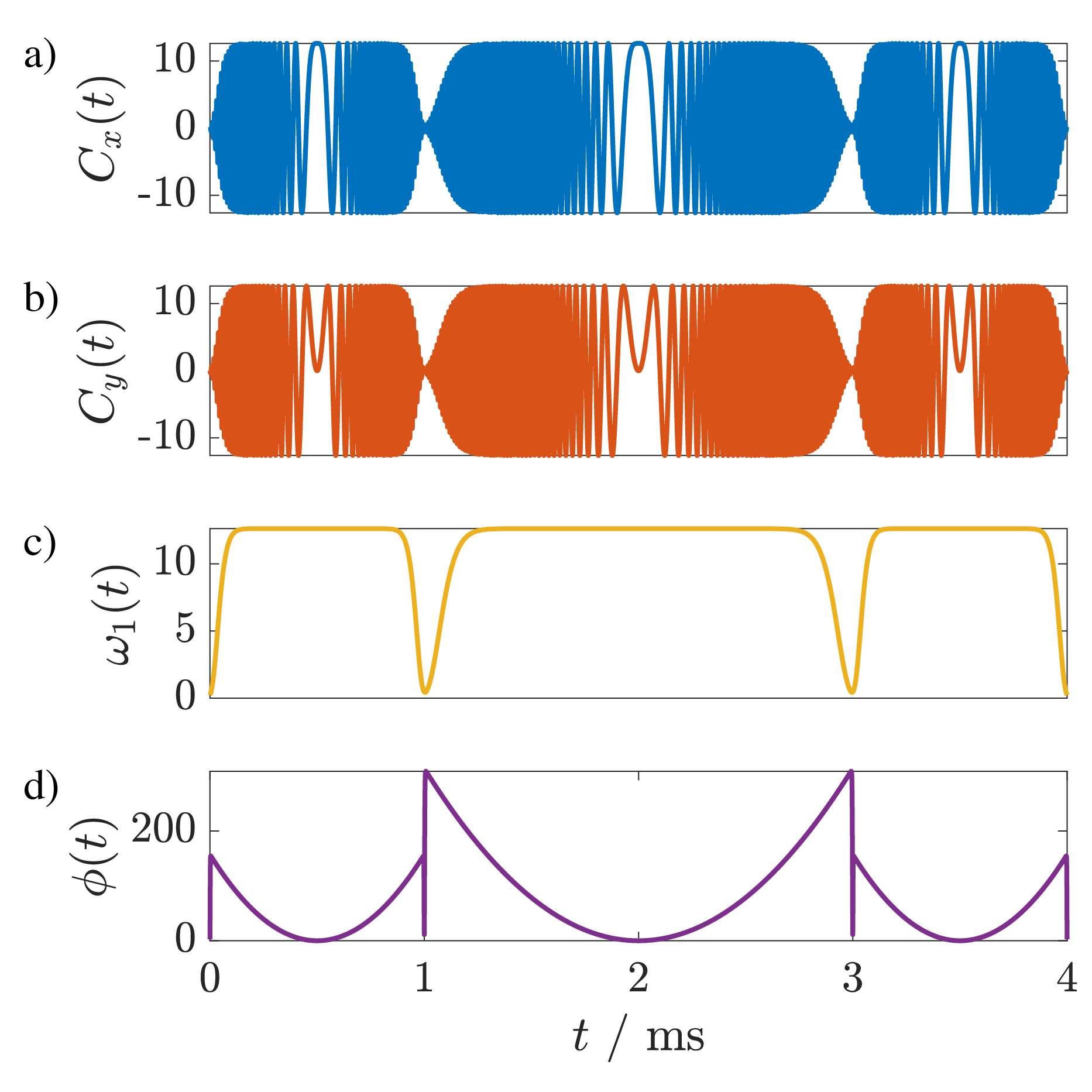}\\
  \caption{Graphical representation of Cartesian ($C_{x}(t)$ and $C_{y}(t)$), and polar ($\omega_{1}(t)$ and $\phi(t)$) components of the composite refocusing chirped sequence, with parameters of \cref{eq:compchirpparam}. The unit for $y$-axis in (a), (b), and (c) is kHz and in (d) is radian.}
  \label{fig:fig_4}
\end{figure}

Including these parameters in \cref{eq:cxcycomp} results in the sequence shown in \cref{fig:fig_4}.

Now we can solve \cref{eq:wnode} for this system giving access to the solution of the unitary propagator for the whole sequence. This result can be incorporated in \cref{eq:rhogamma} and \cref{eq:gamma2col} to have access to the time evolution of the density matrix for a single spin during this three-pulse sequence. For this example we consider the initial state of spins to be $\left| y \right>$, i.e. in \cref{eq:rhogamma} $[c_{1}(0), c_{2}(0), c_{3}(0)]$ is $[0, 1, 0]$.

\begin{figure}
  \includegraphics[width=\linewidth]{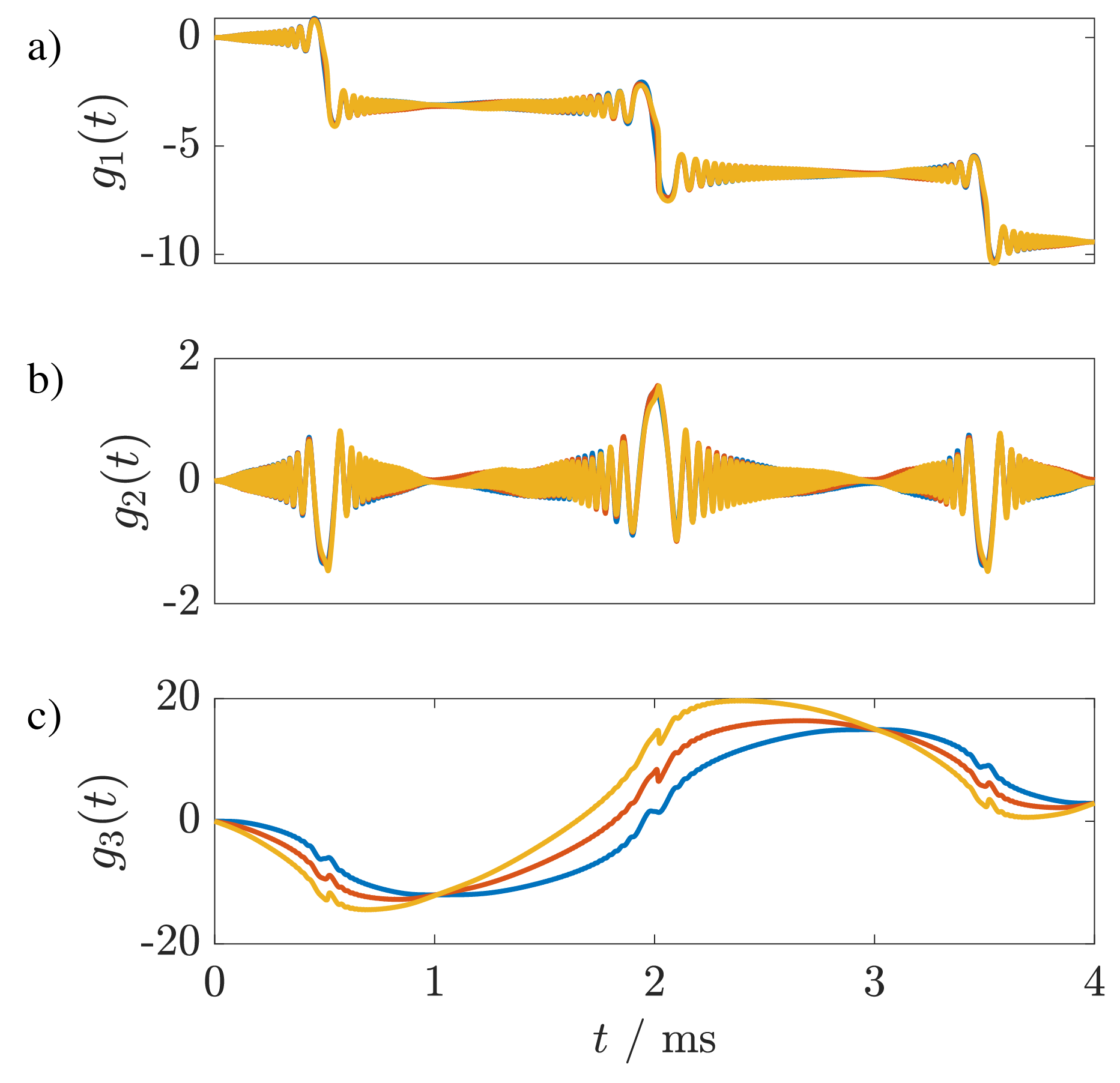}\\
  \caption{Graphical representation of the time-variation of coefficients $g_1(t)$,  $g_2(t)$, and  $g_3(t)$ in \cref{eq:wnode} during the composite refocusing chirped sequence, with parameters of \cref{eq:compchirpparam}.}
  \label{fig:fig_5}
\end{figure}

\begin{figure}
  \includegraphics[width=\linewidth]{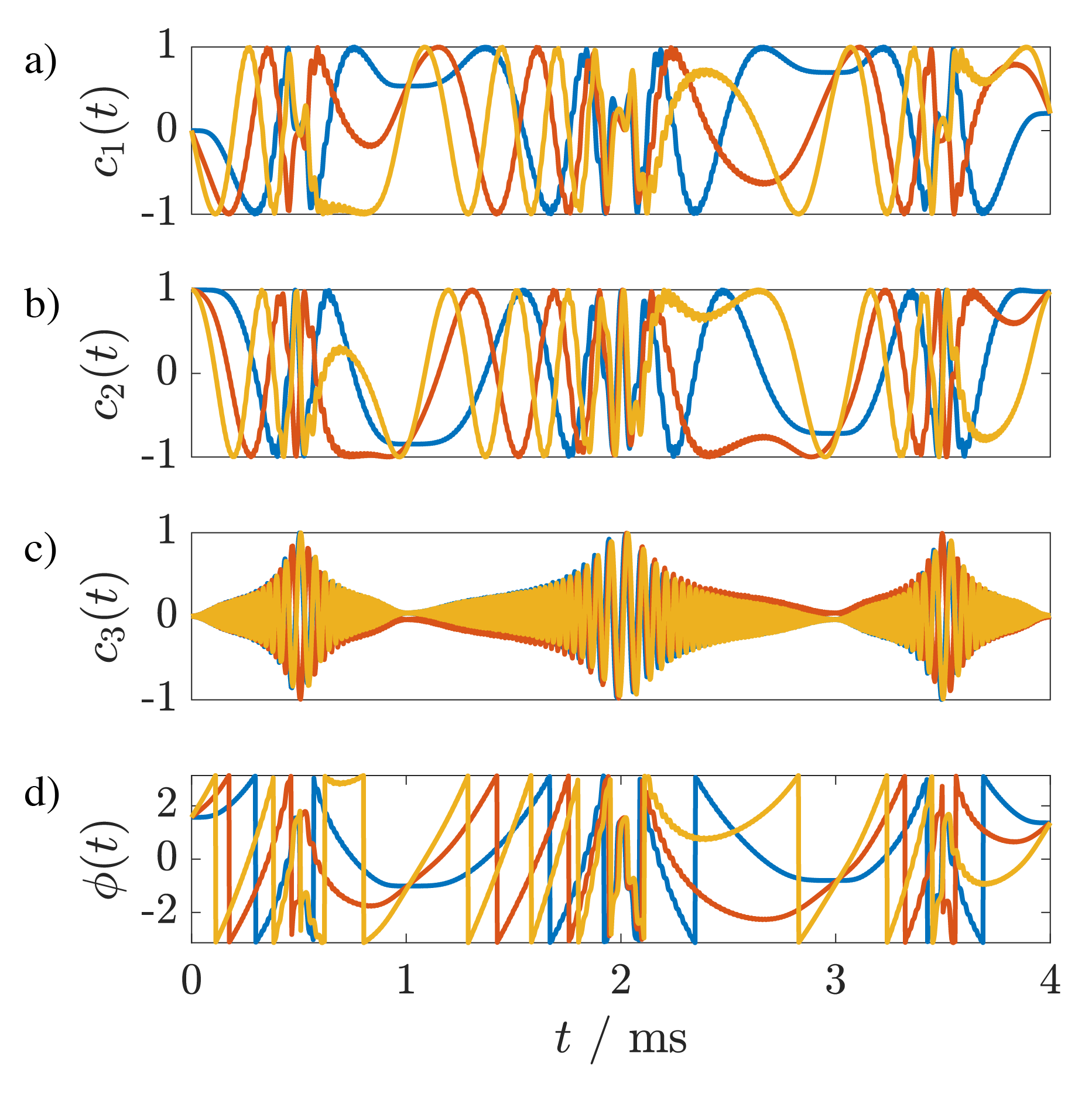}\\
  \caption{Graphical representation of the time-variation of coefficients $c_1(t)$,  $c_2(t)$, $c_3(t)$ in \cref{eq:rhogamma}, and the phases $\phi(t)=\arctan\left(\nicefrac{c_2(t)}{c_1(t)}\right)$ during the composite refocusing chirped sequence, with parameters of \cref{eq:compchirpparam}.}
  \label{fig:fig_6}
\end{figure}

\Cref{fig:fig_5} shows the solutions for $g_{1}(t)$, $g_{2}(t)$, and $g_{3}(t)$ in \cref{eq:wnode} and therefore $\mathscr{U}(t)$ in \cref{eq:usol}. \Cref{fig:fig_6} shows the time evolution of elements of the density matrix in \cref{eq:rhoc} for three different offsets during the pulse sequence of \cref{fig:fig_4}. It is evident that the sequence gives perfect refocusing for signals although they undergo different trajectories during the sequence.

In \cref{subsec:chorus} an application of this formalism will be presented for broadband excitation using the CHORUS sequence.


\section{Some applications}
\label{sec:applic}


\subsection{Broadband homonuclear decoupling}
\label{subsec:psyche}

In this section, we use the approach presented in \cref{subsec:ex2spin} to describe and visualise the spin dynamics for a system of two coupled spin-$\frac{1}{2}$s during the PSYCHE pulse element \cite{Foroozandeh2014,Foroozandeh2014a,Foroozandeh2015,Foroozandeh2016,Foroozandeh2018}. The best way to understand what makes a homonuclear decoupling element is to consider what we want to achieve using this pulse element and then relate those desired and undesired terms to specific members of the basis set. The main property of a J-refocusing pulse element, as in the PSYCHE experiment, as opposed to a conventional inversion/refocusing pulse, is that it acts on both types of interactions, chemical shift and J-coupling. Therefore in the context of a well-known spin-echo experiment, a homonuclear decoupling element is a refocusing (time-reversal) pulse for all interactions in the Hamiltonian. This is the key to distinguishing between the interactions of spins with the main magnetic field (chemical shifts) and interactions of spins with each other (scalar couplings). If we put this in the context of the basis set \cref{eq:basis2} and considering the relevant CTP selection is applied, all terms before and after the PSYCHE pulse element can be written as:

\begin{equation}
\label{eq:saltinittarg}
\begin{dcases}
     P^{+}\longmapsto P^{-}\\
     Q^{+}\longmapsto Q^{-}\\
     P^{+}Q_{z}\longmapsto P^{-}Q_{z}\\
     P_{z}Q^{+}\longmapsto P_{z}Q^{-}
\end{dcases}
\end{equation}

Note that the effect of the PSYCHE pulse element can be seen as a collection of one-to-one maps between members of the basis set, in other words, \cref{eq:saltinittarg} means that for each individual input (existing term before the pulse element) there is a unique output (emerging observable after the pulse element), and anything else generated during the pulse element should be suppressed. This assumption on the mechanism of the PSYCHE pulse element equips us with a useful approach to describe and visualise the spin dynamics during the PSYCHE pulse element. The PSYCHE method takes advantage of a combination of low flip angle double-sweep chirped (saltire) pulse and pulsed field gradient in order to achieve the objectives of \cref{eq:saltinittarg}. In this section, we consider the spin dynamics of a system of two spin-$\frac{1}{2}$s interacting via scalar $J$ coupling.

\begin{figure}
  \includegraphics[width=\linewidth]{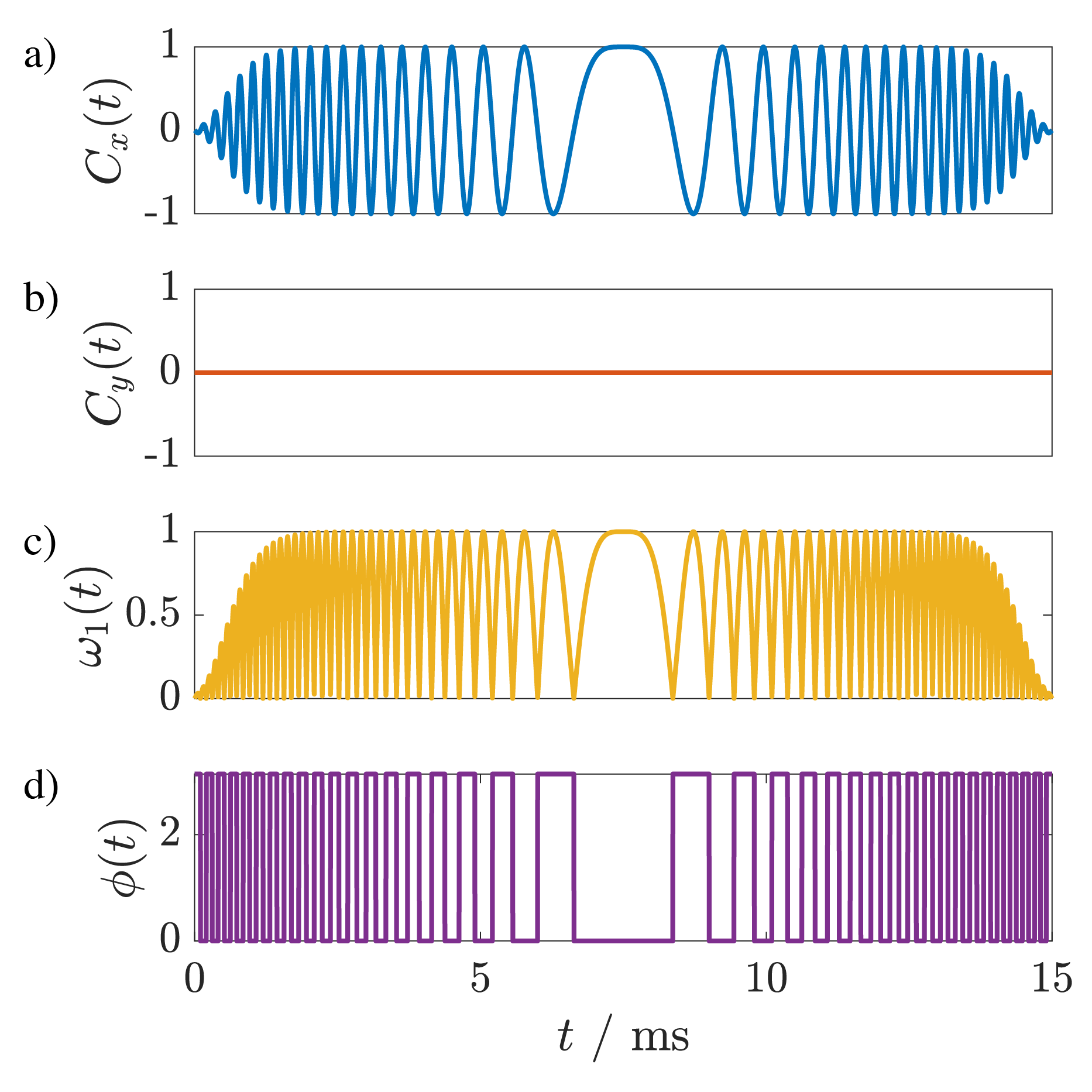}\\
  \caption{Graphical representation of Cartesian ($C_{x}(t)$ and $C_{y}(t)$), and polar ($\omega_{1}(t)$ and $\phi(t)$) components of a saltire pulse with $\tau_{p}=15$ ms, and $\Delta F = 10$ kHz.}
  \label{fig:fig_7}
\end{figure}

A saltire pulse is an average of two counter‐sweeping unidirectional chirped pulses, that is, it is a pulse which simultaneously sweeps in frequency linearly in opposite directions over a frequency range $\Delta F$ in time $\tau_{p}$. As shown in \cref{fig:fig_7}, this averaging changes the pulse from phase‐modulated to amplitude modulated, modifying the appearance of the amplitude envelope. Note that for a saltire pulse:

\begin{equation*}
\delta_{t} = \frac{\tau_{p}}{2}, \qquad \phi_{0}=0, \qquad \delta_{f}=0, \qquad t \in [0,\tau_{p}]
\end{equation*}

The general form of \Cref{eq:suchi} for a saltire pulse can be written as:

\begin{equation*}
S(t)=\frac{1}{2}\sum_{i=1}^{2}\omega_{1} \exp \left[-2^{n+2} \left(\frac{t-\frac{\tau_{p}}{2}}{\tau_{p}}\right)^{n}+\mathrm{i} \left(\frac{(-1)^{i} \pi \Delta F (t-\frac{\tau_{p}}{2})^{2}}{\tau_{p}}\right)\right]
\end{equation*}

which turns the pulse into an amplitude-modulated pulse:

\begin{equation}
S(t)=\omega_{1} \exp \left[-2^{n+2} \left(\frac{t-\frac{\tau_{p}}{2}}{\tau_{p}}\right)^{n}\right] \cos \left[\frac{\pi \Delta F \left(t-\frac{\tau_{p}}{2}\right)^{2}}{\tau_{p}}\right]
\end{equation}

The Hamiltonian for two coupled spin-$\frac{1}{2}$s in the presence of a saltire chirp pulse and a simultaneous pulsed field gradient can be written as:

\begin{equation}
\begin{aligned}
\mathscr{H}(t)=&\Omega_{P} P_{z}+\Omega_{Q} Q_{z} \\
&+\pi J\left(P^{-} Q^{+}+P^{+} Q^{-}+2 P_{z} Q_{z}\right) \\
&+\beta \left(P^{+}+Q^{+}+P^{-} +Q^{-}\right)\\
&+\Omega_{g}(z) \left(P_{z} +Q_{z}\right)
\end{aligned}
\end{equation}

where 

\begin{equation}
\beta = \frac{1}{2} \omega_{1} \exp \left[-2^{n+2} \left(\frac{t-\delta_{t}}{\tau_{p}}\right)^{n}\right] \cos \left[\frac{\pi \Delta F \left(t-\frac{\tau_{p}}{2}\right)^{2}}{\tau_{p}}\right]
\end{equation}

and $\Omega_{g}(z)$ is a frequency offset induced by the pulsed field gradient as in \cref{eq:swg}.

Again, similar to the zero quantum suppression, in order to avoid unnecessary signal loss due to diffusion or excessive spatial encoding we adjust the strength of gradient $G(z)$ so that $\Omega_{g}(z)$ varies in the range $-\pi \Delta F$ to $+\pi \Delta F$.

Finally, the array of the Hamiltonian coefficients for the PSYCHE pulse element can be written as:

\begin{equation}
\label{eq:coeff2salt}
\begin{aligned}
\bm{\gamma}= & \left[ \Omega_{P}+\Omega_{g}(z),\Omega_{Q}+\Omega_{g}(z), \pi J, \pi J, \pi J, \right.\\
& \sqrt{2} \beta, \sqrt{2} \beta, \sqrt{2} \beta, \sqrt{2} \beta, 0, 0, 0, 0, \\
& \left. 0, 0 \right]
\end{aligned}
\end{equation}

Now including these in \cref{eq:ode2col} we can set a system of ODEs for this system. We need to replace $\Omega_{P}$ and $\Omega_{Q}$ with $\Omega_{P} + \Omega_{g}(z)$ and $\Omega_{Q} + \Omega_{g}(z)$ respectively. Additionally we have:

\begin{equation*}
\mathcal{J}=\pi J, \quad \mathcal{B} =\sqrt{2}\beta, \quad \Sigma = \Omega_{P} + \Omega_{Q} + 2\Omega_{g}(z), \quad \Delta = \Omega_{P} - \Omega_{Q}
\end{equation*}

Moreover as the saltire pulse is an amplitude modulated pulse (has only the real part), $\mathcal{B} = \mathcal{B}^{*}$ and matrix $\Gamma$ in \cref{eq: Xmat} is not hermitian, but symmetric.
Additionally, as a result of the pulse being amplitude-modulated the $\mathcal{RF}$ dependence of the flip-angle ($\alpha$) is not asymptotic and can be obtained by the integration of the time-envelope of the saltire pulse:

\begin{equation}
\mathcal{RF}_{max}=\left(\frac{\alpha}{360}\right)\sqrt{\frac{2 \Delta F}{\tau_{p}}}
\end{equation}

In order to see the mechanism of PSYCHE experiment it is crucial to see trajectories of different elements of the basis set and observe how desired terms are prepared for detection and undesired paths are suppressed. For this we need to know the status of the system before the saltire pulse. Since there is only one $90^{\circ}$ pulse before the saltire pulse, some $180^{\circ}$ pulses, and some arbitrary delay we can consider that before the PSYCHE pulse element we have only single quantum terms which can be in the form of a mixture of in-phase (e.g. $P^{+}$) and anti-phase (e.g. $P^{+} Q_{z}$) terms. Of course, during the low-amplitude saltire pulse all other coherence orders will be generated. The idea here is to see how the combination of a low-flip angle saltire pulse and a pulsed field gradient results in an efficient filtration and attenuation of unwanted terms and achieving objectives in \cref{eq:saltinittarg}. 

\begin{figure}
  \includegraphics[width=\linewidth]{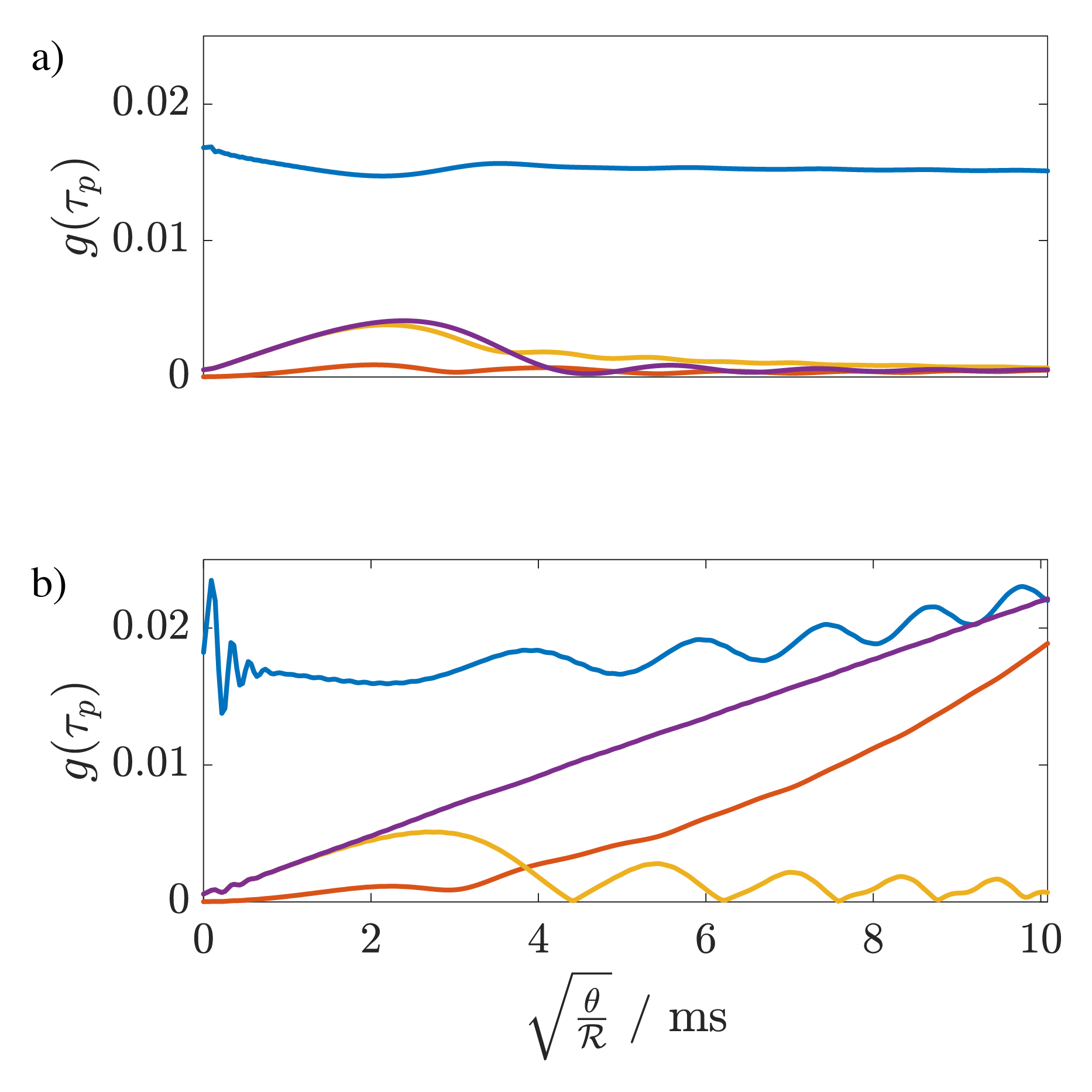}\\
  \caption{Graphical representation of 4 terms $g_{6}(\tau_p)$ (blue),  $g_{8}(\tau_p)$ (red), $g_{10}(\tau_p)$ (orange), and $g_{12}(\tau_p)$ (purple), corresponding to terms on the right side of \cref{eq:saltinittarg}, $P^{-}$, $Q^{-}$, $P^{-}Q_{z}$, and $P_{z}Q^{-}$ respectively as in the basis set of \cref{eq:basis2} as a function of $\xi$ (\cref{eq:xisweep}); a) in the presence, and b) in the absence of the pulsed field gradient. In both cases $\alpha=15^{\circ}$.}
  \label{fig:fig_8}
\end{figure}

Without loss of generality, let us consider a case when we feed the PSYCHE pulse element, consisting of a saltire pulse and a simultaneous pulsed field gradient, with $P^{+}$ only, i.e. at the beginning of the event $g_{7}(0)=1$ and all other coefficients are 0. By looking at \cref{eq:saltinittarg} it is clear that the desired output after pulse element should be $P^{-}$, i.e. we want to maximise $g_{6}(\tau_{p})$ while suppressing all other terms as much as possible. This case study is carried out by considering two independent characteristics of the pulse element: spatiotemporal averaging and low-flip angle excitation. Note that due to the CTP selection and detection of observable terms with negative coherence order we are not interested in single quantum terms with positive coherences at the end of the pulse, $P^{+}$, $Q^{+}$, $P^{+}Q_{z}$, and $P_{z}Q^{+}$, corresponding to $g_{7}(\tau_p)$, $g_{9}(\tau_p)$, $g_{11}(\tau_p)$, and $g_{13}(\tau_p)$ coefficients respectively. The total objectives of the experiment therefore are: i) maximising $P^{-}$ corresponding to $g_{6}(\tau_p)$ as the desired output; ii) minimising other observable (single quantum) terms ($Q^{-}$, $P^{-}Q_{z}$, and $P_{z}Q^{-}$) corresponding to $g_{8}(\tau_p)$, $g_{10}(\tau_p)$, $g_{12}(\tau_p)$, and iii) suppression of terms with zero quantum coherences ($P_{z}$, $Q_{z}$, $P^{-}Q^{+}$, $P^{+}Q^{-}$, and $P_{z}Q_{z}$) corresponding to $g_{1}(\tau_p)$ to $g_{5}(\tau_p)$ respectively and double quantum coherences ($P^{-}Q^{-}$ and $P^{+}Q^{+}$) corresponding to $g_{14}(\tau_p)$ and $g_{15}(\tau_p)$. \Cref{fig:fig_8} shows the variation of these 4 terms at the end of a saltire pulse in the presence and absence of a pulsed field gradient as a function of $\xi$ (\cref{eq:xisweep}).

\begin{figure}
  \includegraphics[width=\linewidth]{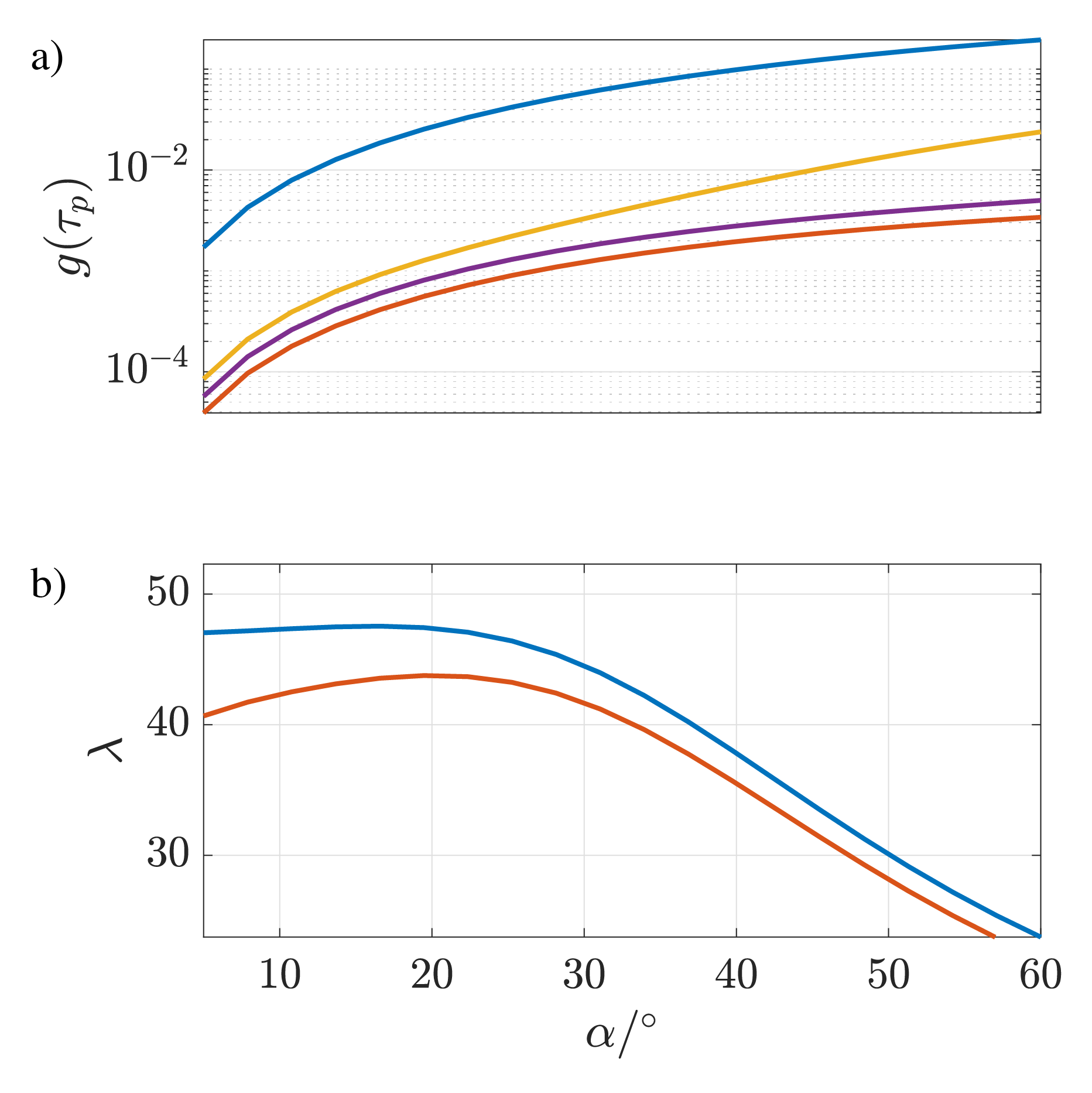}\\
  \caption{a) Graphical representation of the intensities of 4 terms $g_{6}(t)$ (blue),  $g_{8}(t)$ (red), $g_{10}(t)$ (orange), and $g_{12}(t)$ (purple), corresponds to terms on the right side of \cref{eq:saltinittarg}, $P^{-}$, $Q^{-}$, $P^{-}Q_{z}$, and $P_{z}Q^{-}$ respectively as in the basis set of \cref{eq:basis2} versus flip angle ($\alpha$), b) $\lambda$ (\cref{eq:lambdaflip}) with (red) and without (blue) the penalty term; $\tau_{p}=30$ ms, $\Delta F=10$ kHz, $\Delta = 200$ Hz , and $J=10$ Hz.}
  \label{fig:fig_9}
\end{figure}

\Cref{fig:fig_9}(a) shows the flip angle dependence of these 4 terms. It is evident that although the ratio of wanted term $P^{-}$ to unwanted terms ($Q^{-}$, $P^{-}Q_{z}$, and $P_{z}Q^{-}$) is much higher for smaller flip angles, but on the downside, the magnitude of the wanted signal will be significantly smaller. Therefore, it is sensible to seek a compromise for the best suppression of unwanted signals and the best possible sensitivity of the desired signal. For this purpose let us define a parameter $\lambda$ as:

\begin{equation}
\label{eq:lambdaflip}
\lambda=3 \frac{g_{6}(\tau_p)}{\left\|a\right\|}, \qquad a = \left[g_{8}(\tau_p), g_{10}(\tau_p), g_{12}(\tau_p)\right]
\end{equation}

The blue graph in \cref{fig:fig_9}(b) shows the value of parameter $\lambda$ as a function of flip angle. Of course we are interested in a flip angle that maximises both $\lambda$ and $g_{6}(\tau_p)$, therefore we can rewrite the $\lambda$ as:

\begin{equation}
\label{eq:lambdaflip2}
\lambda=3 \frac{g_{6}(\tau_p)}{\left\|a\right\|}+\ln{(g_{6}(\tau_p))}
\end{equation}

Here the term $\ln{(g_{6}(\tau_p))}$ acts as a penalty term to avoid unnecessary signal loss. The red graph in \cref{fig:fig_9}(b) represents the variation of the modified $\lambda$ as a function of flip angle $\alpha$, which shows an optimal flip angle of $\sim 20^{\circ}$. Of course, in \cref{eq:lambdaflip2} data was considered noiseless and only signal-to-artefact ratio was taken into account. In reality with noisy data one can use larger flip angle, as long as the magnitude of unwanted signals remains within the noise variance.

\begin{figure*}
  \includegraphics[width=\linewidth]{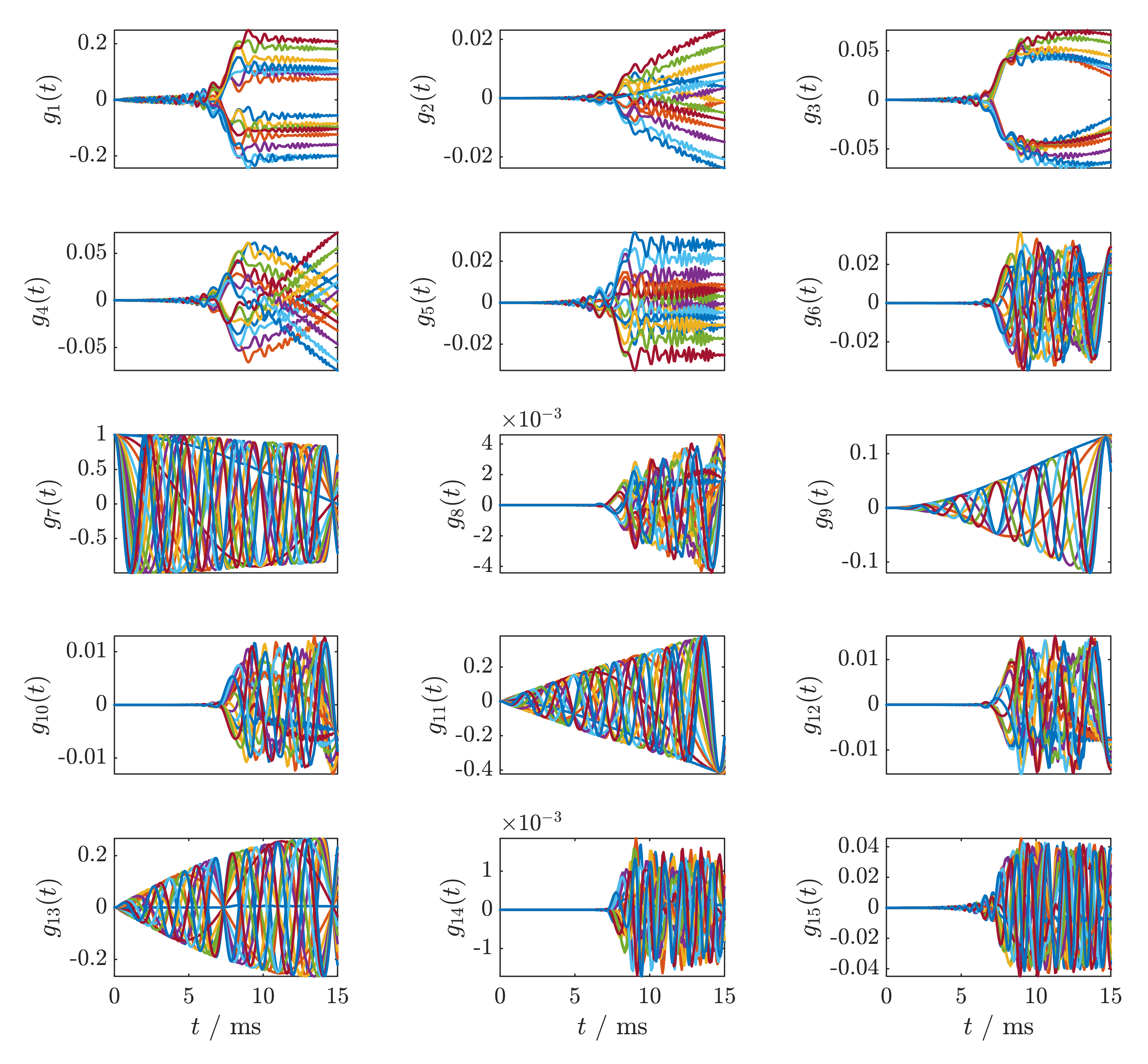}\\
  \caption{Graphical representation of time-dependent coefficients $g_{1}(t)$ to $g_{15}(t)$ during a PSYCHE pulse element; different colours represent time-evolution of terms at different parts of the active volume of the sample.}
  \label{fig:fig_10}
\end{figure*}

\Cref{fig:fig_10} shows time-variation of all 15 coefficients of the members of the basis set in \cref{eq:basis2} at different parts of the active volume of the sample during a PSYCHE pulse element. The idea of phase variation leading to efficient spatiotemporal averaging is obvious from insets $g_{1}(t)$ to $g_{5}(t)$ in \cref{fig:fig_10}. On the other hand, it leads to non-zero averaging for $g_{6}(\tau_p)$, $g_{8}(\tau_p)$, $g_{10}(\tau_p)$, $g_{12}(\tau_p)$.

\subsection{Broadband excitation}
\label{subsec:chorus}

In this section, we use the approach presented in \cref{subsec:exwnla} to describe and visualise the spin dynamics of a group of non-interacting spin-$\frac{1}{2}$s during the CHORUS \cite{Power2016,Power2016a,Foroozandeh2019} pulse sequence. The whole sequence can be written as a sum of three chirped waveforms as in \cref{eq:genchirp,eq:genchirpx,eq:genchirpy}. Note that all three pulses have the same bandwidth $\Delta F$, and for all of them $\phi_{0}=0$ and $\delta_{f} = 0$, but duration, $\tau_{p}$,  position, $\delta_{t}$, and $\mathcal{RF}$ amplitude, $\omega_{1}$, are different.

The total pulse sequence can therefore be written as:

\begin{equation}
C(t)=\sum_{i=1}^{3}\omega_{1}^{(i)} \exp \left[-2^{n+2} \left(\frac{t-\delta_{t}^{(i)}}{\tau_{p}^{(i)}}\right)^{n}+\mathrm{i} \left(\frac{\pi \Delta F (t-\delta_{t}^{(i)})^{2}}{\tau_{p}^{(i)}}\right)\right]
\end{equation}

The Hamiltonian of a single spin-$\frac{1}{2}$ under CHORUS sequence is:

\begin{equation}
\label{eq:hamchorus}
\mathscr{H}(t)=\Omega  \sigma_{z} +C_{x}(t) \sigma_{x}+C_{y}(t) \sigma_{y}
\end{equation}

where

\begin{equation*}
\begin{aligned}
&C_{x}(t)=\sum_{i=1}^{3}\omega_{1}^{(i)} \exp \left[-2^{n+2} \left(\frac{t-\delta_{t}^{(i)}}{\tau_{p}^{(i)}}\right)^{n}\right] \cos\left[\left(\frac{\pi \Delta F (t-\delta_{t}^{(i)})^{2}}{\tau_{p}^{(i)}}\right)\right]\\
&C_{y}(t)=\sum_{i=1}^{3}\omega_{1}^{(i)} \exp \left[-2^{n+2} \left(\frac{t-\delta_{t}^{(i)}}{\tau_{p}^{(i)}}\right)^{n}\right] \sin\left[\left(\frac{\pi \Delta F (t-\delta_{t}^{(i)})^{2}}{\tau_{p}^{(i)}}\right)\right]
\end{aligned}
\end{equation*}

The relative timings for the CHORUS sequence are unique: $\tau_{p}^{(2)} = \nicefrac{\tau_{p}^{(1)}}{2} + \tau_{p}^{(3)}$ with a delay between two $180^{\circ}$ pulses equal to $\nicefrac{\tau_{p}^{(1)}}{2}$. Therefore for the full CHORUS sequence $t \in [0, 2(\tau_{p}^{(1)}+\tau_{p}^{(3)})]$. For example, if the durations of the first ($90^{\circ}$) and the last ($180^{\circ}$) chirped pulses are 500 $\mu$s and 1 ms respectively, the total sequence will be 3 ms long. 

\begin{figure}
  \includegraphics[width=\linewidth]{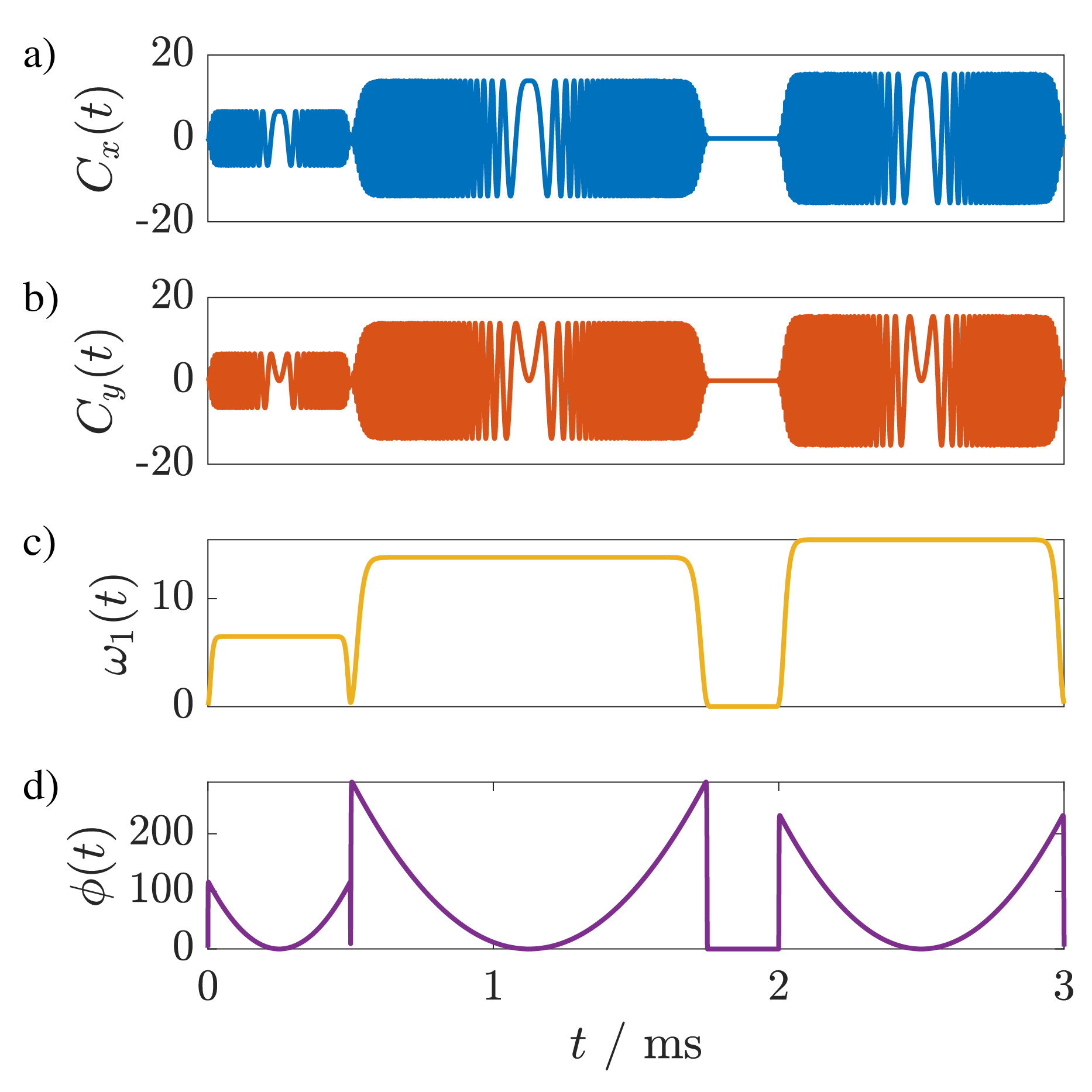}\\
  \caption{Graphical representation of Cartesian ($C_{x}(t)$ and $C_{y}(t)$), and polar ($\omega_{1}(t)$ and $\phi(t)$) components of the composite refocusing chirped sequence, with parameters of \cref{eq:chorusseqparam}.}
  \label{fig:fig_11}
\end{figure}

Here we use Wei-Norman Lie algebra as presented in \cref{subsec:wnla}. We know the general form of matrix $\Xi$ (\cref{eq:xibig}) and its inverse, which is independent of the coefficients of the basis set in the Hamiltonian. Now we can solve \cref{eq:wnode} for this system giving access to the solution of the unitary propagator for the whole sequence. This result can be incorporated in \cref{eq:rhogamma} and \cref{eq:gamma2col} to have access to the time evolution of the density matrix for single spin during CHORUS sequence. 

An example set of parameters for the CHORUS sequence shown in \cref{fig:fig_11} is written as:

\begin{equation}
\label{eq:chorusseqparam}
\begin{dcases}
     \bm{\omega}_{1} = [40787, 86832, 97081] \quad (\nicefrac{\text{rad}}{\text{s}})\\
     \bm{\Delta F} = [300000, 300000, 300000] \quad (\text{Hz})\\
     \bm{\tau}_{p} = [0.0005, 0.00125, 0.001] \quad (\text{s})\\
     \bm{\phi}_{0} = [0, 0, 0] \quad (\text{rad})\\
     \bm{\delta}_{t} = [0.00025, 0.001125, 0.0025] \quad (\text{s})\\
     \bm{\delta}_{f}  = [0, 0, 0] \quad (\text{Hz})\\
\end{dcases}
\end{equation}

\begin{figure}
  \includegraphics[width=\linewidth]{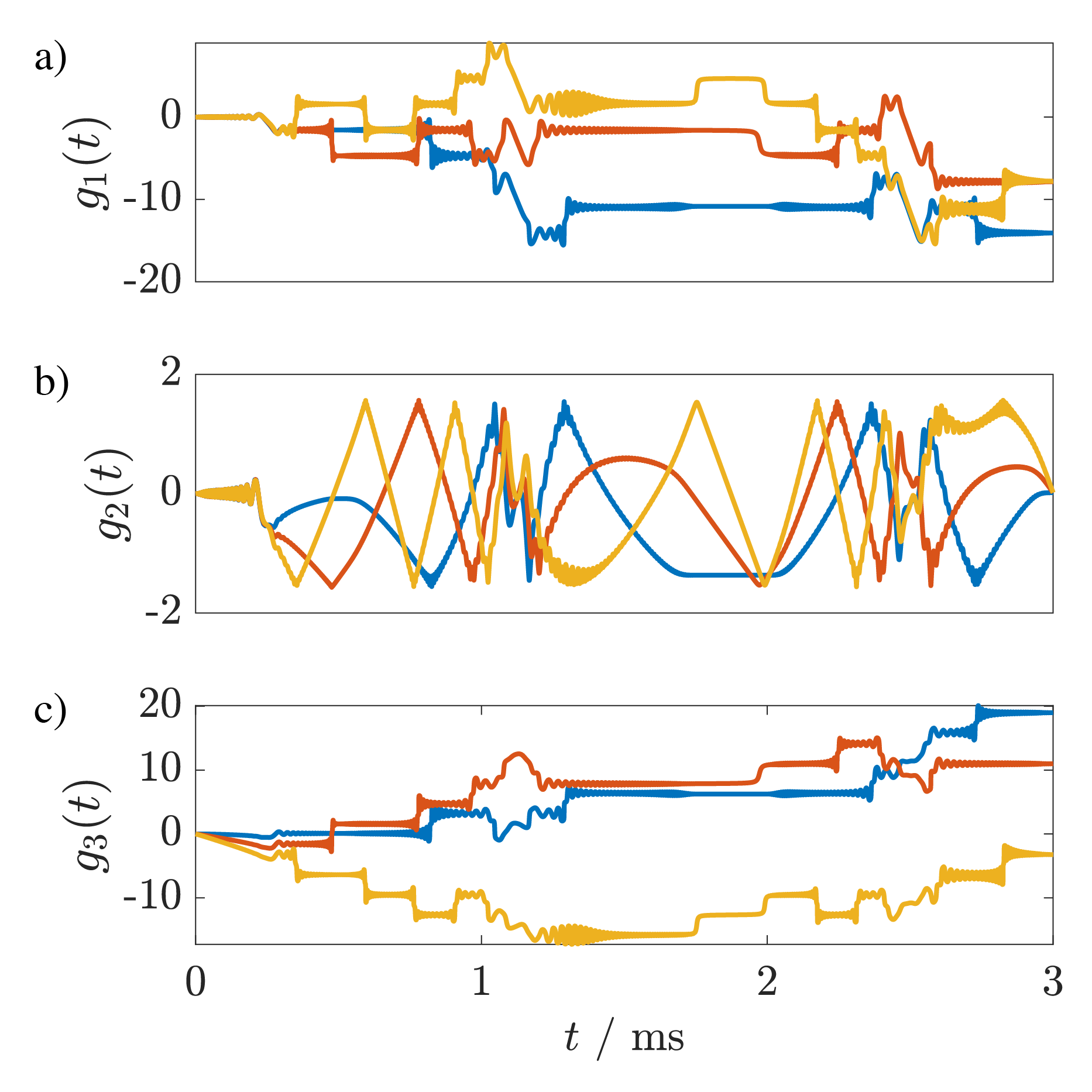}\\
  \caption{Graphical representation of the time-variation of coefficients $g_1(t)$,  $g_2(t)$, and  $g_3(t)$ in \cref{eq:wnode} during the CHORUS sequence, with parameters of \cref{eq:compchirpparam}.}
  \label{fig:fig_12}
\end{figure}

\begin{figure}
  \includegraphics[width=\linewidth]{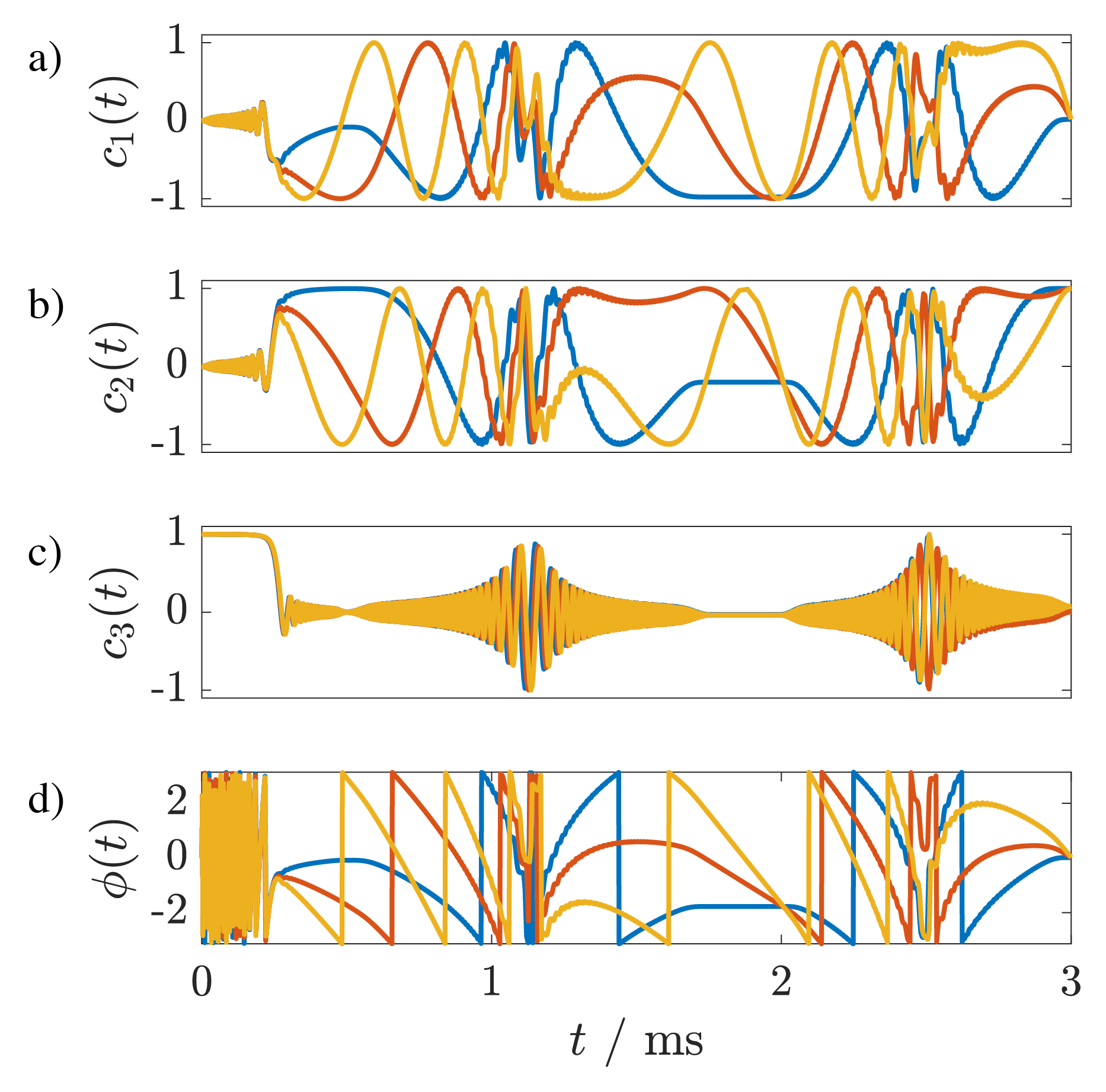}\\
  \caption{Graphical representation of the time-variation of coefficients $c_1(t)$,  $c_2(t)$, $c_3(t)$ in \cref{eq:rhogamma}, and the phases $\phi(t)=\arctan\left(\nicefrac{c_2(t)}{c_1(t)}\right)$ during the CHORUS sequence, with parameters of \cref{eq:compchirpparam}.}
  \label{fig:fig_13}
\end{figure}

\Cref{fig:fig_12} shows the solution of \cref{eq:wnode} for $g_1(t)$,  $g_2(t)$, and  $g_3(t)$, and \cref{fig:fig_13} shows the time-variation of coefficients $c_1(t)$,  $c_2(t)$, $c_3(t)$ in \cref{eq:rhogamma}, and the phases $\phi(t)=\arctan\left(\nicefrac{c_2(t)}{c_1(t)}\right)$ during the CHORUS sequence. 

One remaining issue here is the presence of some residual phase roll across the spectrum. In previous works \cite{Power2016,Foroozandeh2019} it has been proposed that this problem can be fixed by fitting a polynomial function to the phase response of the sequence and add that function directly to the phase of the first pulse ($90^{\circ}$). Here, due to the additional degree of freedom of the generalised chirped pulse, one can approach and solve this problem more efficiently by tweaking the pulse parameters. First, let us write $C_{x}(t)$ and $C_{y}(t)$ for the CHORUS sequence in their most general forms.

\begin{equation}
\label{eq:chorchirpx}
\begin{split}
C_{x}(t)=\sum_{i=1}^{3} & \omega_{1}^{(i)} \exp \left[-2^{n+2} \left(\frac{t-\delta_{t}^{(i)}}{\tau_{p}^{(i)}}\right)^{n} \right] \\
&\cos\left[\phi_{0}^{(i)}+\frac{\pi \Delta F (t-\delta_{t}^{(i)})^{2}}{\tau_{p}^{(i)}}-2 \pi \delta_{f}^{(i)} (t-\delta_{t}^{(i)})\right]
\end{split}
\end{equation}

and

\begin{equation}
\label{eq:chorchirpy}
\begin{split}
C_{y}(t)=\sum_{i=1}^{3} & \omega_{1}^{(i)} \exp \left[-2^{n+2} \left(\frac{t-\delta_{t}^{(i)}}{\tau_{p}^{(i)}}\right)^{n} \right] \\
&\sin\left[\phi_{0}^{(i)}+\frac{\pi \Delta F (t-\delta_{t}^{(i)})^{2}}{\tau_{p}^{(i)}}-2 \pi \delta_{f}^{(i)} (t-\delta_{t}^{(i)})\right]
\end{split}
\end{equation}

It can be seen that the variation of overall phase $\phi_{0}$ and frequency offset $\delta_{f}$ of the $90^{\circ}$ pulse in this 3-pulse sequence can lead to a controllable variation of signal phase across the spectrum. This additional degree of freedom can be harnessed to apply non-linear phase correction instead of the method described in \cite{Power2016,Foroozandeh2019} relying on polynomial fitting. 

The whole approach can be written as a simple optimisation as:

\begin{equation}
\operatorname*{argmin}_{\phi_{0}, \delta_{f}} \Var\left[\phi(\Omega)\right]
\end{equation}

where $\Var$ represents the variance and $\phi(\Omega)$ is the phase of signals at the end of the sequence as a function of the resonance offset $\Omega$. 

\begin{figure}
  \includegraphics[width=\linewidth]{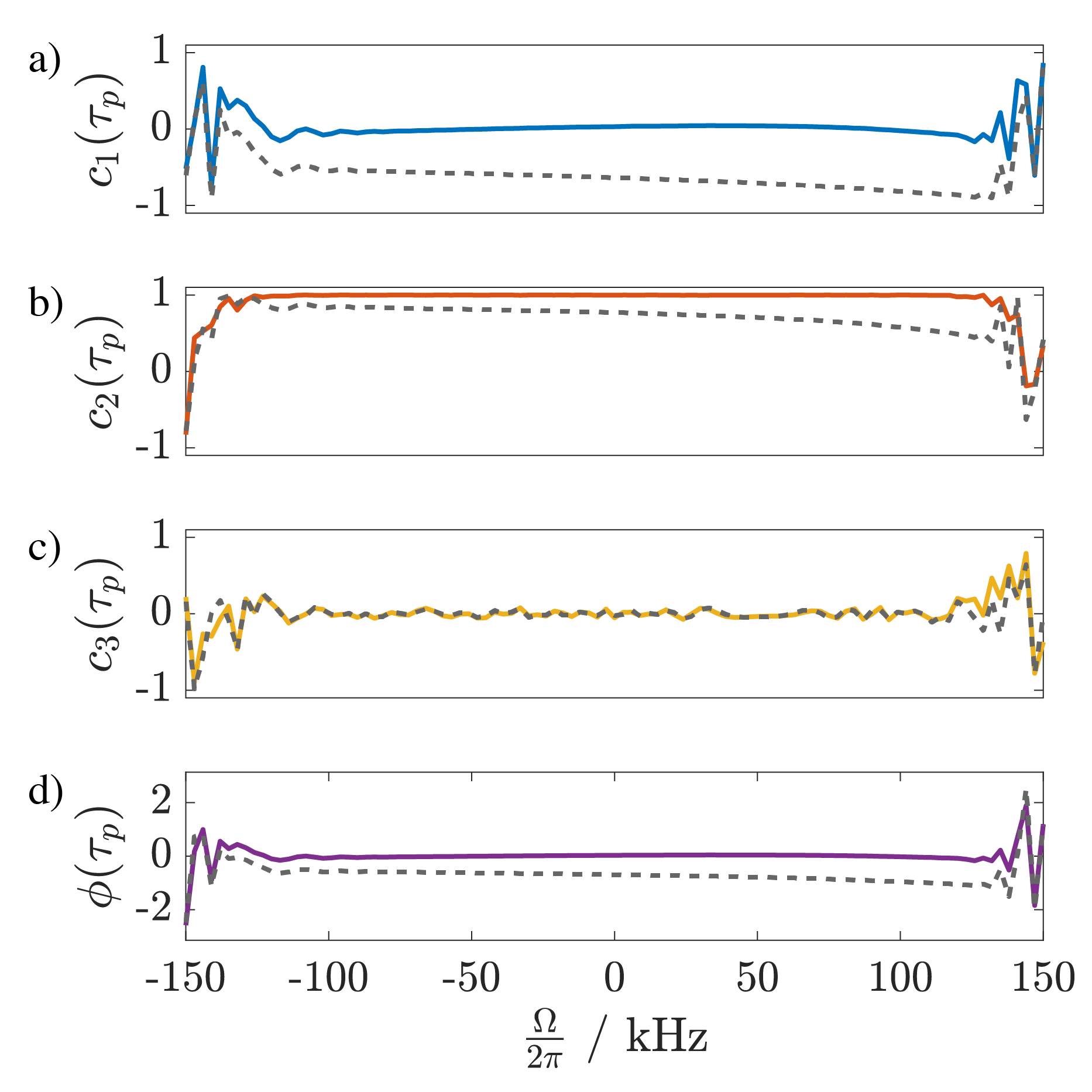}\\
  \caption{Graphical representation of the excitation profile the CHORUS pulse sequence; $c_1(\tau_{p})$,  $c_2(\tau_{p})$, $c_3(\tau_{p})$ in \cref{eq:rhogamma}, and the phases $\phi(\tau_{p})=\arctan\left(\nicefrac{c_2(\tau_{p})}{c_1(\tau_{p})}\right)$ before and after the correction with pulse sequence parameters of \cref{eq:chorusseqparam} and \cref{eq:chorusseqparamopt} respectively.}
  \label{fig:fig_14}
\end{figure}

\Cref{fig:fig_14} shows excitation profiles of CHORUS pulse sequence by plotting coefficients of the density matrix at the end of the sequence i.e. $c_1(\tau_{p})$,  $c_2(\tau_{p})$, $c_3(\tau_{p}), \quad \tau_{p} = 2 \left(\tau_{p}^{(1)} +\tau_{p}^{(3)}\right)$, as a function of the resonance offset $\Omega$.

\begin{figure}
  \includegraphics[width=\linewidth]{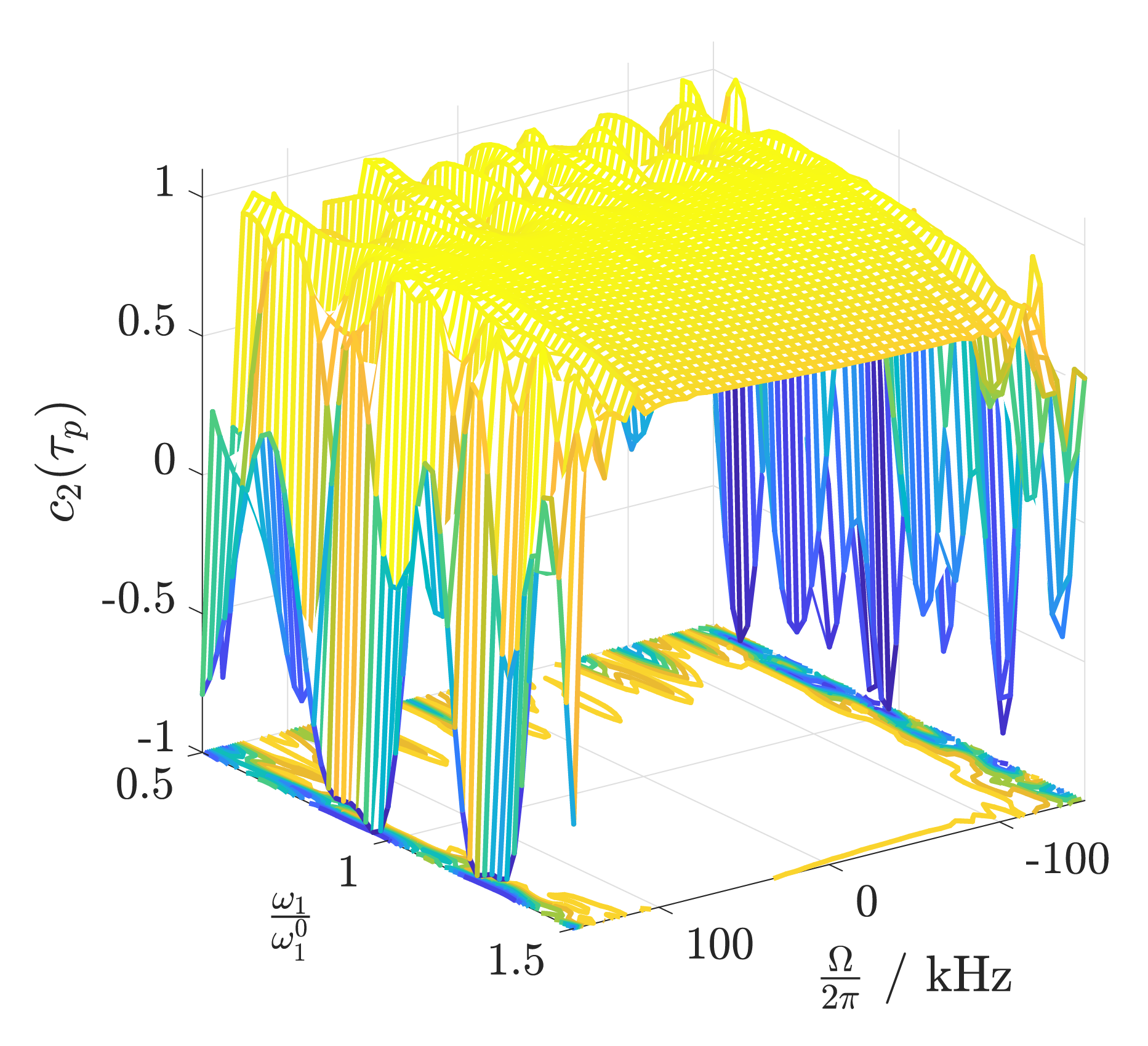}\\
  \caption{3D plots showing the $y$-magnetization at the end of the CHORUS sequence ($c_{2}(\tau_{p})$) as a function of resonance offset ($\Omega$) and relative $\mathcal{RF}$ amplitude $\left(\nicefrac{\omega_{1}}{\omega_{1}^{0}}\right)$.}
  \label{fig:fig_15}
\end{figure}

The set of sequence parameters of \cref{eq:chorusseqparam} after tweaking $\bm{\phi}_{0}$ and $\bm{\delta}_{f}$ is:

\begin{equation}
\label{eq:chorusseqparamopt}
\begin{dcases}
     \bm{\omega}_{1} = [40787, 86832, 97081] \quad (\nicefrac{\text{rad}}{\text{s}})\\
     \bm{\Delta F} = [300000, 300000, 300000] \quad (\text{Hz})\\
     \bm{\tau}_{p} = [0.0005, 0.00125, 0.001] \quad (\text{s})\\
     \bm{\phi}_{0} = [-0.0668, 0.1878, -0.1404] \quad (\text{rad})\\
     \bm{\delta}_{t} = [0.00025, 0.001125, 0.0025] \quad (\text{s})\\
     \bm{\delta}_{f}  = [-200, 0, 0] \quad (\text{Hz})\\
\end{dcases}
\end{equation}

Finally, \cref{eq:wnode} can be solved for a range of $\mathcal{RF}$ amplitudes in order to assess the insensitivity of the CHORUS pulse sequence to variations of the $B_{1}$ field. \Cref{fig:fig_15} shows the magnitude of excited signal as functions of resonance offset and variation of nominal $\mathcal{RF}$ amplitude, using a set of parameters as in \cref{eq:chorusseqparamopt} .

\section{Conclusion}

In the present work, a generalised expression for chirped pulses was introduced that allows pulse sequences consisting of multiple chirped pulses to be written as a single time-continuous waveform. Two mathematical approaches based on Liouville–von Neumann equation and Wei-Norman Lie algebra were presented to calculate and visualise spin dynamics during chirped pulses. Full mathematical treatments were presented on comprehensive examples, and numerical calculations via solutions of ordinary differential equations were presented for broadband homonuclear decoupling using the PSYCHE method and broadband excitation using the CHORUS sequence. Methods proposed in this Perspective have the potential to find applications in NMR, ESR, MRI, and $in$ $vivo$ MRS, where design, simulation, and optimisation of experiments and optimal control of spin systems via parametrised waveforms like swept-frequency pulses are desired. 

\section*{Acknowledgements}
I am grateful to the Royal Society for a University Research Fellowship and a University Research Fellow Enhancement Award (grant numbers URF\textbackslash R1\textbackslash180233 and RGF\textbackslash EA\textbackslash181018). I wish to thank Dr Jean-Nicolas Dumez, Jean-Baptiste Verstraete, and Jonathan Yong for useful comments on the manuscript.

\biboptions{numbers,sort&compress}
\bibliographystyle{elsarticle-num}  
\bibliography{ms}

\end{document}